\pdfoutput=1
\documentclass[sigconf]{acmart}

\newcommand{\method}{\text{MTFM}}

\usepackage{balance}
\usepackage{enumitem}
\usepackage{multirow}
\usepackage{booktabs}
\usepackage{tabularx}
\usepackage{graphicx}
\usepackage{subcaption}
\AtBeginDocument{%
  }

\setcopyright{acmlicensed}
\copyrightyear{2018}
\acmYear{2018}
\acmDOI{XXXXXXX.XXXXXXX}
\acmConference[Conference acronym 'XX]{Make sure to enter the correct
  conference title from your rights confirmation email}{June 03--05,
  2018}{Woodstock, NY}
\acmISBN{978-1-4503-XXXX-X/2018/06}

\usepackage{xcolor}
\usepackage{soul}
\usepackage{comment}
\usepackage{amssymb}

\setuldepth{cat}

\newcommand{\note}[1]{{\color{magenta}$\square$}}

\begin{document}

% \title{MTFM: Foundation Model with Hybrid Target Attention for Industrial Recommendation in Meituan}
\title{MTFM: A Scalable and Alignment-free Foundation Model for Industrial Recommendation in Meituan}


\author{Xin Song$^*$, Zhilin Guan$^*$, Ruidong Han$^{*\dagger}$, Binghao Tang, Tianwen Chen, Bing Li, Zihao Li, Han Zhang, Fei Jiang$^{\dagger}$, Qing Wang, Zikang Xu, Fengyi Li, Chunzhen Jing, Lei Yu, Wei Lin}

\affiliation{%
\{songxin21, guanzhilin02, hanruidong, tangbinghao, chentianwen04, libing65, lizihao30, zhanghan56, jiangfei05, wangqing24, xuzikang02, lifengyi06, jingchunzhen, yulei37, linwei31\}@meituan.com
\\
  \institution{Meituan}
  \city{Beijing}
  \country{China}}

\thanks{* These authors contributed equally to this work.}
\thanks{$\dagger$ Corresponding author.}


\renewcommand{\shortauthors}{MTFM Team}

\begin{abstract}
% 真实的工业级推荐系统往往包含多个场景，在过去Cross-Domain Recommendation以及Multi-scenarios Recommendation一直是推荐系统所研究的重要问题。然而，这些方法大多不仅面对海量的数据需要更多的训练资源，而且这些方法需要对齐不同domain的信息输入，进一步限制了此类方法的推广。
% Industrial recommendation systems typically span multiple scenarios, where cross-domain (CDR) and multi-scenario recommendation (MSR) have been extensively studied.
% However, most of these methods require not only prohibitive training resources when dealing with massive volumes of data, but also strict input alignment across different domains, which further limits the extensibility of such methods. 
Industrial recommendation systems typically involve multiple scenarios, yet existing cross-domain (CDR) and multi-scenario (MSR) methods often require prohibitive resources and strict input alignment, limiting their extensibility.
%In this paper, 我们设计了一个新颖被称之为MTFM的框架来解决上述问题（Meituan Foundation Model）。
We propose \textbf{\method{}} (\textbf{M}ei\textbf{t}uan \textbf{F}oundation \textbf{M}odel for Recommendation), a transformer-based framework that addresses these challenges.
% 不同与DLRM，MTFM基于类似transfomer架构进行建模。它允许不做任何输入层面的对齐，直接将不同领域的数据编码成异构Token，采用统一模型同时建模multi-sceneario知识。
% Unlike DLRM, \method{} is built upon a transformer-like architecture.
% Instead of pre-aligning inputs, it directly transforms cross-domain data into heterogeneous tokens, enabling the model to capture multi-scenario knowledge simultaneously in an alignment-free manner.
Instead of pre-aligning inputs, MTFM transforms cross-domain data into heterogeneous tokens, capturing multi-scenario knowledge in an alignment-free manner.
% 为了降低基座模型资源开销，首先，对于训练数据，我们采用了user level aggregation with multi-scenario sample极大的减少了样本行数进而提升训练吞吐。其次，对于模型架构，我们通过引入Multi-query attention降低显存使用，同时设计hybrid target attention来降低full attention巨大计算开销。
To enhance efficiency, we first introduce a multi-scenario user-level sample aggregation that significantly enhances training throughput by reducing the total number of instances. 
We further integrate Grouped-Query Attention and a customized Hybrid Target Attention to minimize memory usage and computational complexity.
% To address the heavy resource demands of such a foundation model, we first introduce a multi-scenario user-level sample aggregation that significantly enhances training throughput by reducing the total number of instances. 
% Furthermore, we integrate Group-Query Attention and a customized Hybrid Target Attention mechanism to minimize memory usage and mitigate the high computational complexity of standard self-attention.
% 此外，我们系统的优化了训练框架与推理框架，如消除cpu-gpu blocking、Custom Kernel Developmen等进一步提升了训练、推理吞吐。
Furthermore, we implement various system-level optimizations, such as kernel fusion and the elimination of CPU-GPU blocking, to further enhance both training and inference throughput.
Offline and online experiments validate the effectiveness of MTFM, demonstrating that significant performance gains are achieved by scaling both model capacity and multi-scenario training data.
\end{abstract}

\begin{CCSXML}
<ccs2012>
<concept>
<concept_id>10002951.10003317.10003347.10003350</concept_id>
<concept_desc>Information systems~Recommender systems</concept_desc>
<concept_significance>500</concept_significance>
</concept>
\end{CCSXML}
\ccsdesc[500]{Information systems~Recommender systems}

% \keywords{Do, Not, Us, This, Code, Put, the, Correct, Terms, for,


\keywords{Recommender System; Ranking Model; Foundation Model}
\maketitle

\section{Introduction}

%最近一些研究展现了Scaling Law在推荐系统中的潜力，受大模型发展从单一模态到多模态的启发，将单一场景扩展至多场景是非常重要的，一个统一的基座模型可以更好的发挥数据优势，进一步推高模型的智能上限。
The landscape of Large Language Models (LLM) has been redefined by the era of Foundation Models (FMs), where unified architectures have evolved from unimodal to multimodal \cite{bai2025qwen3vltechnicalreport, yang2025kwai, team2025kimi, fu2025vita, liu2026ministral3}. In parallel, the recommendation community has begun to validate the Scaling Laws \cite{kaplan2020scaling, hoffmann2022training}, primarily focusing on the expansion of model parameters and data volume within isolated scenarios \cite{zhai2024actions,zhu2025rankmixer,zhang2024wukong, wang2025mtgrboost, zhou2025onerectechnicalreport, han2024enhancing}. However, the true potential of a Recommendation Foundation Model remains untapped as long as models are confined to single-scenario silos. Inspired by the success of multimodal FMs, we argue that the next frontier lies in bridging cross-scenario heterogeneity. By extending from single-scene optimization to a unified multi-scenario framework, we can leverage diverse behavioral signals to further unlock the benefits of Scaling Laws, thereby elevating the performance upper bound and establishing a more scalable paradigm for recommendation.

% Recent studies have demonstrated the potential of Scaling Laws within recommendation systems\cite{zhai2024actions,han2025mtgr}. 
% Inspired by the evolution of Large Language Models from unimodal to multimodal\cite{bai2025qwen3vltechnicalreport, yang2025kwai, team2025kimi, fu2025vita, liu2026ministral3}, extending single-scenario frameworks to multi-scenario environments is essential. 
% A unified foundation model can better leverage the data advantages, thereby further elevating the upper bound of model intelligence.

% 一个好的foundation model该具备哪些特点呢？。我们认为一个好的foundation model除了具有好的效果，还应该具有scalability，extensibility，efficiency。其中scalability指通过不断scale up可以取得更好的效果；extensibility指针对多个场景或增加新场景建模是方便容易的；efficiency指面对多业务下数据激增，模型在训练、推理上具有低成本。
To realize this vision, we contend that an ideal Recommendation Foundation Model must embody three quintessential properties: Scalability, Extensibility, and Efficiency. 
First, Scalability requires the model to yield consistent and predictable performance gains as the parameter size and data volume expand. 
Second, Extensibility dictates that the model can seamlessly adapt to an arbitrary number of existing scenarios or integrate emerging ones with low cost. 
Finally, Efficiency; the framework must maintain low computational overhead during both training and inference, even when handling the massive data volumes generated across diverse scenarios.

% 对于scalability，当前主流的方法主要分为transformer based以及非transformer based又或者称之为deep learning recommendation model(DLRM)。对于transformer-based，这种方法会将特征输入拆分成一个个token再使用类transformer架构统一建模。对于DLRM方法，模型会被拆分成多个模块进行建模，此类方法会寻找一个可堆叠的模块进行scale up，而不是scale up整个模型。

% In terms of scalability, current methods are primarily categorized into two paradigms: Transformer-based and non-Transformer-based (often referred to as Deep Learning Recommendation Models (DLRM)). 
% Transformer-based approaches\cite{han2025mtgr,zhai2024actions,deng2025onerec} partition input features into individual tokens and leverage Transformer-like architectures for unified modeling. 
% Conversely, DLRM methods\cite{zhu2025rankmixer,zhang2024wukong} adopt a modular design, decomposing the model into various functional components. 
% Rather than scaling the entire architecture, these methods achieve scalability by scaling up some specific stackable modules.

% 当前主流的多场景建模方法主要依赖于DLRM paradigm, 一个经典的多场景建模方法首先要收集多个领域的数据并针对特征进行极其复杂的处理，来保证不同领域的数据是同构的。对于不同质的部分需要丢弃或者做额外的padding。随后会将模型参数或结构会被拆分成domain-specific以及domain-invariant进行建模来获取更好的效果。
Existing multi-scenario modeling efforts \cite{sheng2021one,shu2024adaptive,zhang2024m3oe,xiaoyu2025soft} primarily follow a "harmonize-then-decompose" paradigm. 
These approaches typically aggregate data from multiple scenarios and enforce input homogeneity with a fixed template.
Any heterogeneous components must be either discarded or handled through an additional padding operator.
Structurally, they rely on disentangling model parameters into domain-invariant and domain-specific components (e.g., via MoE-based architectures \cite{ma2018modeling, fedus2022switch}) to capture shared knowledge while preserving scenario-specific nuances.

%然而此类方法存在多个缺点。
% However, such approaches are constrained by several critical limitations: 
% 1) 严格的homogeneity要求不仅损害了extensibility，还损害了模型效果。不同scenario的特征体系存在较大差异，例如，在外卖业务中，同时存在了food推荐以及store推荐等多种推荐形式，这些场景供给不同，展现形式也不同，因此建模所用的特征差异巨大。面对场景间巨大的差异性，涉及数百甚至上千特征的人工特征对齐极易出错且很难获得持续拓展。强行删除不能对齐的特征也会导致效果损失。
% 2) 模型结构不满足scalaiblity。这类方法往往依赖于专家经验手工设计各种精巧的结构，这些来学习多业务下的共性与区别，这些方法本身可能是次优的，也没有经过scale up的验证。
% 2) Model architectures lack scalability. 
% These methods often rely on handcrafted, intricate structures designed based on expert heuristics to capture the commonalities and differences across various scenarios. 
% Such designs may be inherently suboptimal and have not been rigorously verified through scaling.
% 3）使用更多数据同时带来过高成本。对于DLRM，训练成本与数据量成线性关系，多场景数据的使用会导致高昂的训练成本。
% 3) Leveraging more data entails prohibitive costs. In the DLRM paradigm, training costs scale linearly with the volume of data, meaning that the integration of multi-scenario data inevitably leads to exorbitant training expenses.

However, these conventional paradigms fall short of the aforementioned hallmarks for a foundation model. 
1) Rigidity in Extensibility: Industrial recommendations often encompass highly divergent scenarios with heterogeneous feature spaces. For instance, in our food delivery platform, the feature schemas for restaurant recommendation and food recommendations are inherently distinct due to different types of supply and UI presentations. The reliance on strict feature alignment is unsustainable in industrial ecosystems.
Forcing hundreds of heterogeneous features into a fixed template is not only error-prone but also leads to significant information loss. 
2) Lack of Architectural Scalability: Most existing structures are handcrafted based on expert heuristics to handle a fixed set of scenarios. 
Such intricate, scenario-tailored designs lack the structural flexibility needed to benefit from the scaling laws observed in other scenarios. 
3) Prohibitive Computational Costs: Within the traditional paradigm, training costs increase linearly with data volume. 
The brute-force integration of massive multi-scenario data leads to exorbitant expenses, rendering the model economically inefficient for real-world deployment.

To address these challenges, we present the Meituan Foundation Model (MTFM), a unified framework engineered to satisfy the three quintessential properties. 
% MTFM将不同领域的数据抽象成异构Token，避免了繁琐的输入对齐，极大提升了extensibility。在得到Token输入后，MTFM扩展了具有良好scalability的MTGR架构，采用统一的transformer-like模型对信息进行编码。它不需要复杂的对于模型结构的手动设计，而是通过注意力机制直接保证模型能够在scale up时自动的从数据中学习domain-specific以及domain-invariant的知识。
First, to ensure Extensibility, MTFM moves beyond rigid feature templates by abstracting multi-domain inputs into a unified sequence of heterogeneous tokens. 
This token-based representation allows the model to ingest disparate signals across scenarios without manual alignment or information loss. 
Building upon this, we employ a transformer-inspired backbone designed for Scalability. 
By replacing handcrafted heuristic-based structures with a deep self-attention mechanism, MTFM empowers the model to automatically capture both universal behavioral laws and scenario-specific nuances as the model and data scale.

% 在模型结构上，为了在此基础上进一步降低训练、推理成本，首先我们采用Multi-Query Attention来降低显存开销。其次，受LLM中稀疏注意力应用的启发，提出hybrid target attention，不同与HSTU或者是MTGR全注意力机制，也不同与onerec中全稀疏的lazy decoder，hybrid target attention介于二者中间，将稀疏注意力与全注意力混合，大部分层采用稀疏注意力，仅少部分层采用全注意力。对于MTFM，稀疏注意力进一步退化为Target Attention，即Key与Vale来自于全部Token，而Query来自代表候选的那些Token。

To achieve superior Efficiency without compromising modeling capacity, we introduce Hybrid Target Attention (HTA). 
While full-attention backbones excel in expressiveness, their quadratic complexity is prohibitive for industrial-scale long sequences. 
Inspired by the application of sparse attention in LLMs \cite{yuan2025native,qwen3next,coreteam2026mimov2flashtechnicalreport}, HTA serves as a strategic middle ground: it selectively applies full attention to a few critical layers to maintain global dependency capture, while utilizing target attention for the majority of the stack. 
By further integrating Grouped-Query Attention (GQA) \cite{ainslie2023gqa}, MTFM significantly compresses the memory footprint and improves the training throughput, enabling the processing of more multi-scenario tokens with minimal overhead.

% 我们还做了大量数据pipeline、训练与推理框架上的优化以提升效率。在数据方面，我们采用了多场景下的用户粒度压缩，通过增加token数以大幅降低样本行数。在训练框架上，我们针对cpu-gpu管道优化消除系统中阻塞点，同时针对多个算子进行Kernel fusion进行优化。我们还额外引入了半精度推理、Structured Sparsity等推理优化手段。
% We have also implemented extensive optimizations across the data pipeline, as well as the training and inference frameworks, to enhance overall efficiency. 
% On the data front, we employ user-level aggregation for multi-scenario samples, which significantly increases token density per sequence while drastically reducing the total number of training instances. 
% For the training framework, we streamline the CPU-GPU pipeline to eliminate system bottlenecks and apply kernel fusion across various operators. 
% Furthermore, we incorporate inference optimization techniques such as half-precision (FP16) computation and structured sparsity.

Beyond architectural innovations, we implement a suite of systematic optimizations to bridge the gap between theoretical models and real-world deployment. 
On the data front, we utilize user-level aggregation from multiple scenarios to transform fragmented samples into dense sequences, substantially boosting training throughput.
At the system level, we nearly double both training and inference throughput by eliminating critical bottlenecks through a range of techniques, such as CPU-GPU pipeline orchestration, kernel fusion, and structured sparsity.
These co-designs ensure that MTFM is economically viable for large-scale production.

% 我们设置了不同尺寸的模型验证了MTFM在美团工业级推荐任务上的表现。对于离线，我们的模型在多个业务多个任务上均取得了更好的效果，对于CTR任务，GAUC平均增长0.36pp, 最大增长0.76pp；对于CXR任务，GAUC平均增长0.29pp，最大增长0.53pp。对于在线，我们的模型当前已经部署在部分场景上，在Shenqiangshou(SQS) Coupon-Package Recommendation上取得了订单+3.44%的效果，在Pinhaofan(PHF) Group-buying Recommendation取得了订单+0.40%的效果。
% We evaluated the performance of MTFM across various model scales on industrial-level recommendation tasks at Meituan. 
% In offline evaluations, our model achieved significant average gains of +xx\% in CTR (Click-Through Rate) and +xx\% in CTCVR (Click-Through Conversion Rate) across multiple business lines. 
% Online A/B tests further confirm its superiority; in the coupon-package recommendation task, MTFM delivered a +xx\% increase in CTR and a +xx\% increase in conversion
% volumes.

We evaluate the performance of MTFM through both offline experiments and online A/B testing across various scenarios in Meituan.
Offline experiments demonstrate that MTFM consistently outperforms state-of-the-art baselines across multiple scenarios and objectives. 
Specifically, for the CTR (Click-Through Rate) task, MTFM achieves an average GAUC (Grouped Area Under the Curve) improvement of 0.36pp (percentage points), with a maximum gain of 0.76pp. 
For CTCVR (Click-Through Conversion Rate) tasks, the model yields an average GAUC increase of 0.29pp and a peak improvement of 0.53pp. 
% Beyond offline metrics, MTFM has been deployed in several production environments. 
Furthermore, Real-world online A/B testing shows significant business growth: a +2.98\% increase in orders for the Shenqiangshou (SQS) Coupon-Package Recommendation, and a +1.45\% boost for the Pinhaofan (PHF) Food Recommendation.

%我们的主要贡献如下：
% MTFM扩展了从单一场景到多场景的scale up路径，同时具有scalability, extensibility, and efficiency。
% 我们在标准注意力基础上引入了Hybid target attention, 结合GQA，极大降低了训练，推理开销。
% 我们针对数据pipeline、训练、推理都做了系统性优化来进一步减少资源开销。
% 我们做了详细的实验从离线以及在线上都证明了MTFM的有效性。
Our main contributions are summarized as follows:
\begin{itemize}

\item Unified Foundation Architecture: We propose MTFM, a multi-scenario foundation model that uses heterogeneous tokenization to ensure seamless extensibility and scalability across disparate domains.

\item Hybrid Target Attention: We introduce HTA, a novel attention mechanism combining sparse-dense layers and GQA to achieve an optimal trade-off between modeling capacity and computational efficiency.

\item System-Model Co-design: We implement system-level optimizations to ensure the economic viability of MTFM in industrial-scale production.

\item Empirical Validation: Extensive offline and online evaluations at Meituan demonstrate significant gains, proving MTFM's real-world effectiveness.

\end{itemize}

\section{Related Work}

\subsection{Multi-Scenario Recommendation}
% 一些方法也尝试超越这一paradigm，\cite{zhang2023modeling, zhang2024unified}引入了图的方法将异构信息看作是不同节点进行建模，但是采用图难以捕捉用户与候选的交叉信息。
Existing research primarily follows a "harmonize-then-decompose" paradigm, which aims to balance commonality and diversity by disentangling model parameters into domain-invariant and domain-specific components. For instance, STAR \cite{sheng2021one} pioneered this approach with a star topology that anchors domain-specific parameters around shared centered weights. To further capture complex feature interactions, M$^{3}$OE \cite{zhang2024m3oe} leverages three specialized MoE modules to learn hierarchical user preferences across common, domain, and task aspects. More recently, MLoRA \cite{yang2024mlora} introduces a more parameter-efficient alternative by adapting LoRA modules for individual domains.
% Beyond this dominant paradigm, some studies \cite{zhang2023modeling, zhang2024unified} have explored graph-based modeling to represent heterogeneous information as distinct nodes. Nevertheless, these graph-based approaches often struggle to capture the information sufficiently between users and candidate items.

Moving beyond parameter-level disentanglement, recent studies have increasingly explored multi-scenario foundation models to achieve superior generalization. 
This line of work often adopts a "Foundation-Expert" architecture. For example, ExFM \cite{recommendation2025external} distills knowledge from ultra-large-scale models into scenario-specific experts, while LFM4Ads \cite{zhang2025large} utilizes multi-granularity mechanisms to transfer comprehensive representations across features and tasks. 
Similarly, \cite{li2025realizing} incorporates HSTU as a robust encoder to bridge foundational target information with downstream expert models. 
However, these approaches predominantly rely on a two-stage pipeline rather than a fully end-to-end foundation model, which potentially limits the model's performance upper bound.

\subsection{Ranking Model with Scalability}

Current methods are primarily categorized into two paradigms: Transformer-based and non-Transformer-based (often referred to as Deep Learning Recommendation Models, DLRM).
% Unlike traditional DLRM models, where research hotspots primarily focus on feature interaction  \cite{deepFM,DCNv2,zhang2024wukong} and behavior sequence modeling \cite{DIN, DIEN,SIM,LREA} to enable recommendation ranking models to capture various heterogeneous features (sequential/non-sequential), 
Transformer-based approaches partition input features into individual tokens and leverage Transformer-like architectures for unified modeling. 
Specifically, HSTU \cite{zhai2024actions} reconceptualizes recommendation as a sequential transduction task, employing a high-performance architecture to capture user behavior patterns. 
To further enrich feature interactions, MTGR \cite{han2025mtgr} extends this by incorporating cross-features and bidirectional attention mechanisms. 
Recognizing the computational demands of such models, OneTrans \cite{zhang2025onetrans} addresses inference efficiency via a pyramid-style tail truncation strategy. 
More recently, scaling laws have been explored within generative frameworks \cite{deng2025onerec,zhou2025onerecv2technicalreport}, and these efforts also remain rooted in the Transformer backbone.
Conversely, traditional DLRM adopts a modular design, decomposing the system into distinct functional components. 
These methods achieve scalability by selectively expanding specific stackable modules. 
For instance, RankMixer \cite{zhu2025rankmixer} enhances the classic DLRM architecture by integrating Per-token Feed-Forward Networks to boost model capacity. 
Similarly, WuKong \cite{zhang2024wukong} attains scalability through a specialized design of stacked layers, combining Factorization Machine blocks with Linear Compress blocks to effectively increase model scalability.
\section{Methodology}
% 在本章节中，首先，我们介绍多场景数据组织方式；其次，详细阐述MTFM模型架构，包括异构tokenization模块和混合注意力架构；最后，讨论训练和部署过程中的优化技巧，这些技巧对于将MTFM成功应用于工业级多场景推荐系统至关重要。
In this section, we first introduce the multi-scenario data arrangement method in Section~\ref{subsec:data-arrangement}.
Next, we elaborate on the MTFM model architecture in Section~\ref{subsec:model-arch}, including the heterogeneous tokenization module and hybrid attention architecture.
Finally, we discuss optimization techniques during training and deployment in Section~\ref{subsec:train-deploy-opt}, which are critical for successfully applying MTFM to industrial-scale multi-scenario recommendation systems.
\subsection{User Sample Aggregation with Multi-Scenario Data} \label{subsec:data-arrangement}
%在这个小节，我们将MTGR的数据组织方式扩展到了多场景建模
In this subsection, we extend the data arrangement method from MTGR \cite{han2025mtgr} for multi-scenario modeling.
%, whose overview is shown in Figure~\ref{fig:data}.
% 跟MTGR一样，我们在训练时将样本按照用户粒度聚合，在推理时将候选按照请求粒度聚合，区别是，在MTFM中，聚合的样本包含多场景候选和特征
Following MTGR, we aggregate samples at the user level during training and at the request level during inference. The difference is that in MTFM, the aggregated samples contain candidates and features from multiple scenarios.

% \begin{figure}[t]
%     \centering
%     \includegraphics[width=\linewidth]{figs/DataArrangement.pdf}
%     \caption{Data Arrangement of MTFM}
%     \label{fig:data}
% \end{figure}

% 在训练阶段，我们把一个时间窗口下的所有曝光行为按照用户粒度聚合
For offline training, we aggregate all exposure behaviors in a specific time window from multiple scenarios at the user-level.
% 具体来说，每个用户对应一个训练样本D_u = (X_u, Y_u)，其中X_u和Y_u分别是输入特征和标签
Specifically, each user $u$ corresponds to multiple training samples, denoted by $\mathbb{D}_u = (X_u, Y_u)$, where $X_u$ and $Y_u$ are the sets of input features and labels, respectively. In the following, we omit the subscript
$u$ for brevity when there is no ambiguity.
% 输入和标签的具体定义如下
In more detail, the input feature set and the label set with user-level aggregation are denoted as follows:
\begin{align}
    X &= [\{H_i\}_{i=1}^{N_H}, \{R_i\}_{i=1}^{N_R}, \{U^s\}_{s=1}^{N_S}, \{ \{C_i^s, I_i^s \}_{i=1}^{N_E^s} \}_{s=1}^{N_S} ] \\
    Y &= \{ \{Y_i^s \}_{i=1}^{N_E^s} \}_{s=1}^{N_S},
    H_i = \{h_{ij}\}_{j=1}^{N_H^i},
    R_i = \{r_{ij}\}_{j=1}^{N_R^i}
\end{align}
where $C_i^s, I_i^s$, and $Y_i^s$ are the sets of cross features, item-side features, and labels of the $i$-th exposure in scenario $s$.
$N_E^s$ is the number of exposures in scenario $s$, and $N_S$ is the number of scenarios.
${U^s}$ is the set of scenario-specific user features in scenario $s$, $\{H_i\}$ is the set of historical interaction sequences, and $\{R_i\}$ records the user’s most recent behaviors. Each $h_{ij}$ or $r_{ij}$ represents an action, including multiple features such as item ID, item tag, etc., and $N_H^i, N_R^i$ is the sequence length.
Since the timing of behaviors in $R_i$ may overlap with the sample aggregation window, some actions $r_{ij}$ might occur after the exposure represented by $C_i^s$ and $I_i^s$. 
Modeling these directly introduces the risk of information leakage. 
Following the approach in MTGR, we will address this by constructing a dynamic mask based on interaction timestamps in the next section.
% 所有用户在所有场景的训练样本被用来训练一个统一的MTFM模型
% The set of training samples from all users across all scenarios is used to train MTFM as a unified model.

% 需要注意的是，由于用户在不同场景的行为序列都是用户兴趣的一部分，因此用户历史序列和实时序列, i.e., H_i & R_i, 在所有场景中进行共享。然而，对于用户属性、交叉和物品侧特征，这些特征则是scenario-specific的。这是受场景间的客观区别所致，例如，商家推荐场景天然缺乏商品tag特征，导致这两个场景feature schame天然无法对齐。
Notably, as user behavior sequences across diverse scenarios constitute integral parts of their overall interests, both historical and real-time sequences ($H$ and $R$) are shared across all scenarios. In contrast, for scenario-specific features—including user attributes $U$, cross-features $C$, and item-side characteristics $I$ are treated as scenario-specific. This configuration is dictated by the fundamental structural divergence between domains; for instance, the restaurant recommendation scenario naturally lacks specific food tags, rendering the feature schemas across these scenarios intrinsically incompatible for direct alignment.

For online inference, each scenario deploys its corresponding subgraph in MTFM and aggregates the features of all candidates at the request level. Our subgraph restricts computation to components relevant to the current scenario, bypassing scenario-specific modules belonging to other scenarios.
Specically, the input feature for scenario $s$ is denoted by $X^s = [U^s, \{H_i\}_{i=1}^{N_H}, \{R_i\}_{i=1}^{N_R}, \{C_i^s, I_i^s \}_{i=1}^{N_C^s} ]$, where $N_C^s$ is the number of candidates in one request of scenario $s$.

\begin{figure*}[t]
    \centering
    \includegraphics[width=\linewidth]{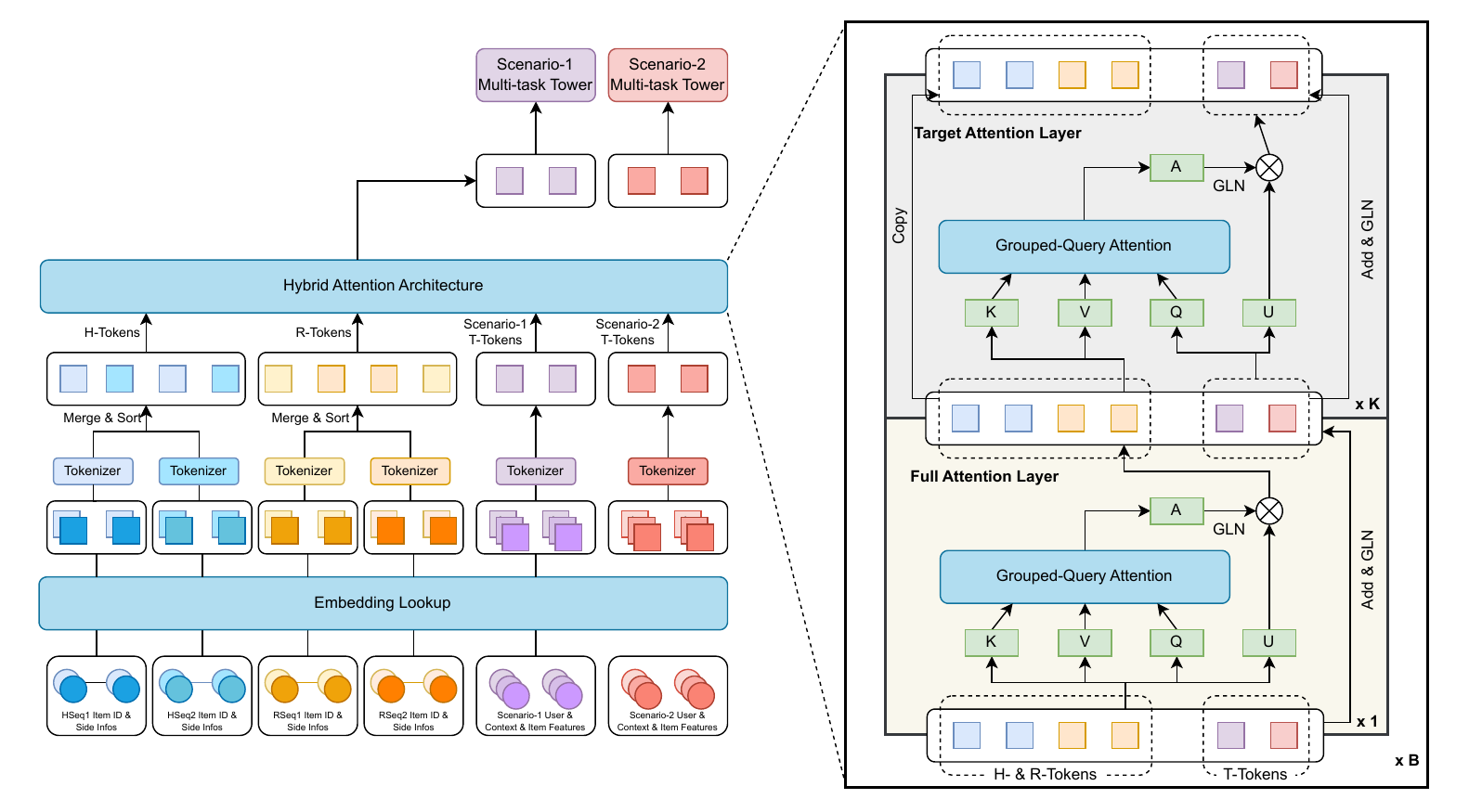}
    \caption{The Model Architecture of MTFM}
    \label{fig:model}
\end{figure*}

\subsection{MTFM Model Architecture} \label{subsec:model-arch}
In this subsection, we introduce the model architecture of MTFM as shown in Figure~\ref{fig:model}.

\subsubsection{Heterogeneous Tokenization} \label{subsubsec:hetero-token}
% 在这个小节，我们介绍将输入特征表示成一个变长且异构的token序列的方法。
We first detail how our model represents the input features as a variable-length sequence of heterogeneous tokens.

% 我们的模型有三大类token，包括H-token，R-token，和T-token
There are three primary types of tokens in our model, including H-token, R-token, and T-token.
% 每个在用户历史序列的item对应一个H-token，先把item特征embedding化，然后用一个MLP把embedding映射到一个统一的维度d_model，从而得到它的token embedding
Taking $\{ H_i\}$ as an example, each item $h_{ij}$ in $\{ H_i\}$ corresponds to an H-token, whose embedding is generated by first embedding raw item features and subsequently applying an MLP to project them into a unified dimension $d_{\text{model}}$:
\begin{equation}
    \mathbf{h}_{ij} = \text{MLP}_i(\text{Emb}(h_{ij}))
\end{equation}
% 由于不同历史序列所包含的特征存在差异，因此其对应的H-token采用不同的MLP作为tokenizer。
Different $\text{MLP}_{i}$ are employed as tokenizers for the H-tokens corresponding to different historical sequences, due to the heterogeneity of features contained within these sequences.

% 之后，我们把所有历史序列的token按照时间顺序排序，得到一个timestamp-aware的embedding矩阵 H
After that, we sort the token embeddings of items in all historical sequences in chronological order to form a new embedding matrix $\mathbf{H} \in \mathbb{R}^{L_H \times d_{\text{model}}}$, where $L_H$ is the total number of items in all historical sequences.

% 类似地，我们把所有实时序列tokenize到一个embedding matrix R
Similarly, the real-time sequences are tokenized into an embedding matrix $\mathbf{R} \in \mathbb{R}^{L_R \times d_{\text{model}}}$, where $L_R$ is the total number of items in all real-time sequences.

% 每个曝光行为对应一个T-token，它的输入特征拼接了用户画像、交叉、item侧的特征。具体地，一个曝光行为的token embedding定义如下：
Each exposure behavior corresponds to a T-token, whose input features are the concatenation of the user profile, cross and target item features. Specically, the token embedding of an exposure behavior on a target item is obtained as follows:
\begin{equation}
    \mathbf{t}_{i}^s = \text{MLP}_{s} (\text{Emb}(U^s) \| \text{Emb}(C_i^s) \| \text{Emb}(I_i^s ))
\end{equation}
where $\|$ means column concatenation. All the exposure behaviors are tokenized into an embedding matrix $\mathbf{T} \in \mathbb{R}^{L_T \times d_{\text{model}}}$, where $L_T$ is the total number of exposures in all scenarios.

% 最后，所有H、R、T-tokens堆叠在一起组成混合注意力架构的初始输入
Finally, all the H-, R-, and T-tokens are stacked to form the initial input embedding matrix to our Hybrid Attention Architecture:
\begin{equation}
    \mathbf{X}^{(0)} = (\mathbf{H}; \mathbf{R}; \mathbf{T}) \in \mathbb{R}^{N \times d_{\text{model}}}
\end{equation}
where $(;)$ means row concatenation and $N = L_H+L_R+L_T$ is the total number of tokens which can vary for different users.

\subsubsection{Hybrid Target Attention Architecture} \label{subsubsec:hybrid-target}
While Transformers are natural candidates for modeling the resulting heterogeneous token sequences, their $O(n^2)$ complexity poses significant computational challenges in multi-scenario settings where sequence lengths increase substantially.
To balance modeling efficacy and efficiency, we propose a Hybrid Target Attention architecture. 
Specifically, our model consists of $B$ stacked blocks; to mitigate the quadratic bottleneck, each block interleaves one full attention layer with $K$ efficient Target Attention layers.
This hybrid design allows the model to capture global dependencies while significantly reducing the computational overhead compared to standard Transformers.

\begin{figure}[h]
    \centering
    \includegraphics[width=\linewidth]{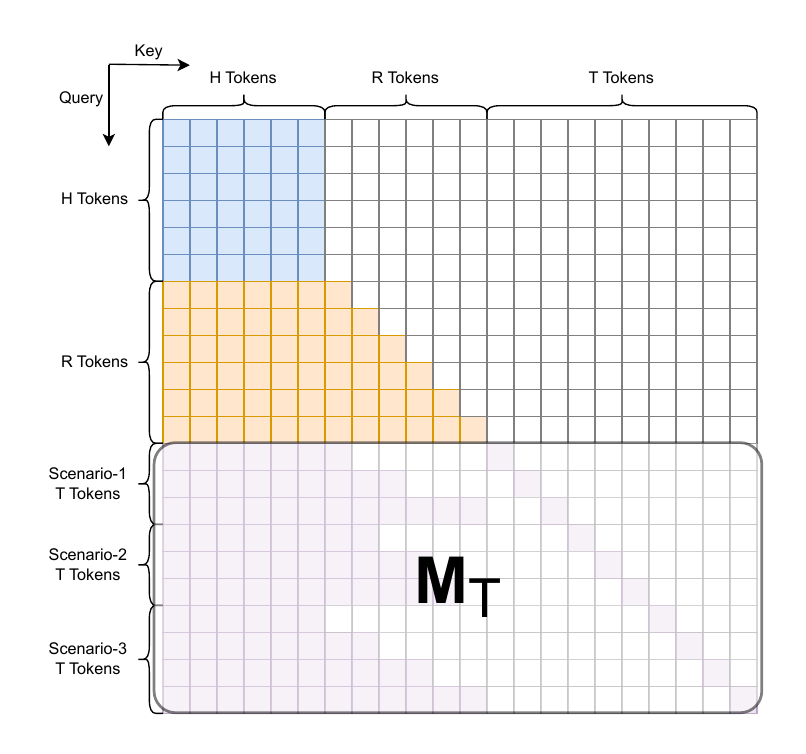}
    \caption{An Example of Dynamic Mask}
    \label{fig:mask}
\end{figure}

We first introduce the Full Attention Layer and the Target Attention Layer, respectively. 
In the Full Attention Layers, we first apply group layer normalization to the input tokens: H-tokens from different historical sequences are assigned to different groups; similarly, R-tokens from different real-time sequences are assigned to different groups; and T-tokens from different scenarios also belong to distinct groups. This design enables the model to better adapt to the diverse distributions arising from different sequences and token features:
\begin{equation}
    \widetilde{\mathbf{X}}^{(l)} = \text{GLN}(\mathbf{X}^{(l)} )
\end{equation}
% 最后，归一化的token embedding被输入HSTU进行序列建模
Finally, the normalized token embeddings are fed into the HSTU for sequence modeling.
% 此外，为了进一步降低计算开销，我们采用的MQA
We adopt Grouped-Query Attention (GQA) to further reduce computational overhead.
% 在第l层的全注意力层可以被正式定义为：
Formally, a full attention layer at the $l$-th layer of our hybrid attention architecture is defined as follows:
\begin{gather}
    \mathbf{U}^{(l)}, \{\mathbf{Q}^{(l,h)}\}_{h=1}^{H}, \{\mathbf{K}^{(l,g)}, \mathbf{V}^{(l,g)} \}_{g=1}^G  = \text{Split}(\phi_1(f_1^{(l)}(\widetilde{\mathbf{X}}^{(l)} ))) \\
    \textbf{A}^{(l,h)} = \phi_2(\mathbf{Q}^{(l,h)}  \mathbf{K}^{(l,g)T}  \odot \mathbf{M}) \mathbf{V}^{(l,g)}, g = \lceil h/r \rceil, \forall h   \\
    \textbf{A}^{(l)} = \textbf{A}^{(l,1)} \| \cdots \| \textbf{A}^{(l,H)} \\
    \mathbf{X}^{(l+1)}  = f_2^{(l)}(\text{GLN}(\mathbf{A}^{(l)} ) \odot \mathbf{U}^{(l)} ) + \mathbf{X}^{(l)} 
\end{gather}
where $H$ is the number of query heads, $G$ is the number of key-value heads, $r = H / G$ is the number of query heads per key-value group; $f_{1}^{(l)}$ and $f_{2}^{(l)}$ denote linear layers; $\phi_1$  and $\phi_2$ are nonlinearity; and $\mathbf{M} \in \mathbb{R}^{N \times N}$ is our dynamic mask matrix to aviod information leakage.
Because it was sorted in advance according to timestamps, we can construct a dynamic mask according to the following rules:
\begin{enumerate}
    \item The H-tokens are visible to all tokens;
    \item The R-tokens are visible only to tokens with later timestamps to avoid information leakage;
    \item The T-tokens are visible only to themselves.
\end{enumerate}
Figure~\ref{fig:mask} illustrates an example of our dynamic mask.

% 在Target Attention Layer中，我们只更新T-token的Embedding，非target token的embedding经过一个shortcut连接传递到下一层
In the Target Attention Layers, we update only the embeddings of T-tokens, while the embeddings of other tokens from the previous layer are passed to the next layer via a shortcut connection.
% 首先，我们从归一化后的token embedding matrix和动态mask中取出T-token对应的部分
First, we extract the sub-matrix corresponding to the T-tokens from the normalized token embedding matrix and the dynamic mask matrix:
\begin{align}
    \widetilde{\mathbf{X}}^{(l)}_T &= \widetilde{\mathbf{X}}^{(l)}[L_H+L_R:] \\
    \mathbf{M}_T &= \mathbf{M}[L_H+L_R:]
\end{align}
Then, we apply HSTU with GQA to update only the embeddings of T-tokens.
% 在第l层的Target注意力层可以被正式定义为：
Formally, a target attention layer at the $l$-th layer of our hybrid attention architecture is defined as follows:
\begin{gather}
    \mathbf{U}^{(l)}_T, \{\mathbf{Q}^{(l,h)}_T\}_{h=1}^H = \text{Split}(\phi_1(f_{uq}^{(l)}((\widetilde{\mathbf{X}}_T^{(l)} ))) \\
     \{\mathbf{K}^{(l,g)}, \mathbf{V}^{(l,g)}\}_{g=1}^G  = \text{Split}(\phi_1(f_{kv}^{(l)}(\widetilde{\mathbf{X}}^{(l)} ))) \\
    \textbf{A}^{(l,h)}_T  = \phi_2(\mathbf{Q}_T^{(l,h)}  \mathbf{K}^{(l,g)T}  \odot \mathbf{M}_T) \mathbf{V}^{(l,g)}, g = \lceil h/r \rceil, \forall h  \\
    \textbf{A}_T^{(l)} = \textbf{A}_T^{(l,1)} \| \cdots \| \textbf{A}_T^{(l,H)} \\
    \mathbf{X}^{(l+1)}_T  = f_2^{(l)}(\text{GLN}(\mathbf{A}_T^{(l)} ) \odot \mathbf{U}_T^{(l)} ) + \mathbf{X}_T^{(l)}
\end{gather}
% 最后，我们将T-token的新embedding与其他token的上一层embedding进行拼接，从而生成target attention层的输出
% 通过这种基于Target Attention的混合架构，我们将计算资源优先分配给处理更关键的T-tokens，在显著降低计算资源消耗的同时确保性能不受影响。
Finally, we concatenate the new embeddings of the T-tokens with the embeddings from the previous layer of other tokens to generate the output of the target attention layer.
\begin{equation}
    \mathbf{X}^{(l+1)} = (\mathbf{X}^{(l)}[:L_H+L_R]; \mathbf{X}^{(l+1)}_T)
\end{equation}

% 最后一层的T-tokens的embedding被输入到一个MMoE，来计算多场景不同目标的预估分
The embeddings of the T-tokens from the final layer are input into an MMoE\cite{ma2018modeling} module to compute prediction scores for various objectives across multiple scenarios.

By employing this hybrid architecture, we reduce the computational complexity from $O(N^2)$ to $O(\frac{KNL_T + N^2}{K+1})$, where $L_T \ll N$. 
% 在我们的实验部分展示了HTA可以获得两倍的训练吞吐且没有任何损失。
Our experiments demonstrate that HTA achieves a 2x throughput speedup in training without compromising model accuracy.

\subsection{Training and Deployment Optimization} \label{subsec:train-deploy-opt}
\label{sec:optimization}
\subsubsection{Data Pipeline of MTFM}
\begin{figure}[t]
    \centering
    \includegraphics[width=\linewidth]{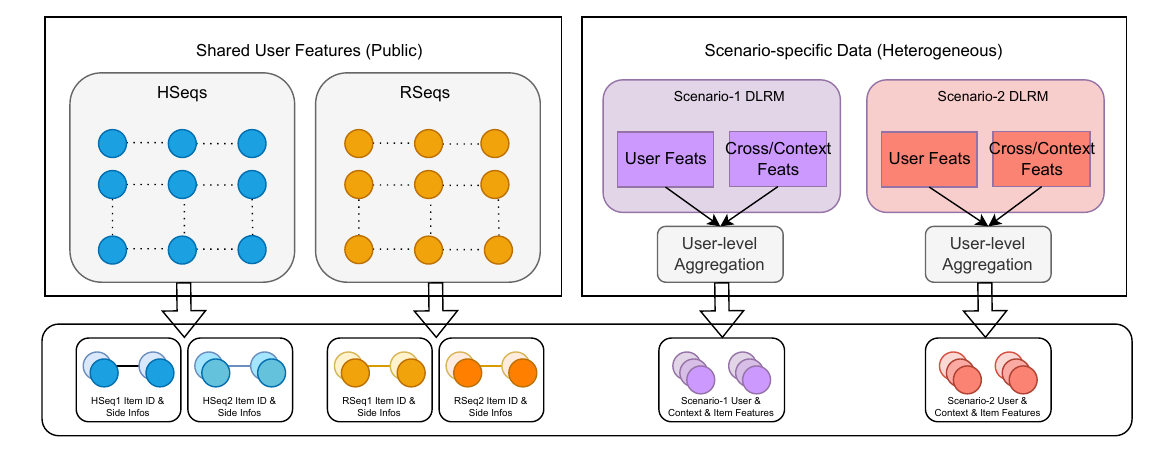}
    \caption{Data Pipeline of MTFM}
    \label{fig:data_pipeline}
    \vspace{-15px}
\end{figure}

% 本节介绍 MTFM 的数据链路设计。与MTGR一样，我们采用了user-level的样本压缩来减少训练数据的行数，并且更进一步的拓展到多场景下的user-level压缩。
% 具体的，我们将特征拆分为全场景共享以及场景独立两个部分，全场景共享特征主要为{H_i}，{R_i},全场景独立的特征主要为{U, C, I}。我们先针对场景独立的特征在每个场景内部按user-level进行聚合。and then concatenated column-wise to construct a unified training sample，最后与全场景共享数据按user-level进行合并，避免了pipeline重复计算和冗余存储，提升了数据链路的鲁棒性和资源利用效率。
In this section, we delineate the data pipeline design of MTFM. Following the precedent set by MTGR, we utilize user-level sample compression to diminish the cardinality of the training set. We further extend this technique into a multi-scenario user-level compression framework, which is shown in Figure~\ref{fig:data_pipeline}.
Specifically, we decouple the feature space into two distinct components: scenario-agnostic (shared) features, such as $H, R$, and scenario-specific features, such as $U, C, I$. We first perform user-level aggregation on the scenario-specific features within their respective local scenarios. These aggregated components are then concatenated column-wise to construct a unified feature representation, which is finally merged with the scenario-agnostic data at the user level. This pipeline effectively circumvents computational redundancy and storage overhead, significantly improving systemic robustness and resource utilization efficiency.
% This section details the data pipeline architecture of MTFM.
% Followed by MTGR, we use a user-level compression to reduce the number of rows in the training data. But we further expand this compression across various scenarios
% To accommodate the modeling requirements of heterogeneous data, we have re-engineered the data pipeline. 
% By adopting a paradigm that integrates shared features with scenario-specific features, we established a unified data processing workflow across all scenarios that are shown in Figure~\ref{fig:data_pipeline}
% 针对不同业务场景下候选目标特征不一致的问题，各业务数据先按 user-level 进行聚合，再将各业务的数据按列拼接，形成统一样本，兼顾了多业务的独特性与一致性。
% To address feature disparities among targets in different scenarios, we implement a user-level aggregation strategy. Data from individual scenarios are first aggregated and then concatenated column-wise to construct a unified training sample. This approach effectively harmonizes the uniqueness of specific scenarios with the consistency required for a unified model.
% 最后与用户全生命周期行为序列及全业务用户实时序列公共特征拼接，避免了pipeline重复计算和冗余存储，提升了数据链路的鲁棒性和资源利用效率。
% Finally, these samples are joined with shared features, including the user's lifelong behavioral sequences and real-time cross-scenario sequences. This design eliminates redundant computations and storage overhead, significantly enhancing the robustness and resource efficiency of the data infrastructure.
\subsubsection{Training Optimization}

% 在LLM领域，通过对文本的tokenization，特征均位于GPU上，训练高效。而在推荐系统中，特征工程繁重并且很多特征位于CPU上,这就导致了位于GPU上的模型和CPU上的特征之间存在阻塞(block)。由于特征处理、模型融合等环节涉及大量Host与Device之间的数据同步和串行依赖,流程中存在较多同步点,Host必须等待Device侧操作完成后才能继续执行,导致整体流程出现较多阻塞(Blocking)。这些CPU-GPU Pipeline Stall严重影响了整个训练性能。
% 为此,我们通过CUDA Profiler系统性地检查了训练框架中的CPU/GPU阻塞问题。优化策略主要包括两个方面:一是消除同步点,使CPU和GPU执行相互掩盖(overlap);二是优化Device端的内存操作,减少Device-to-Device(D2D)拷贝开销。 具体而言,我们针对频繁的张量索引赋值操作进行了D2D优化,将原本需要多次D2D拷贝的操作合并为单次原子操作。通过系统性地消除Pipeline阻塞点并优化D2D操作,最终实现了40%的训练吞吐提升。
% 针对Flash Attention 2在自定义稀疏掩码场景下的性能瓶颈，我们提出了基于异步拷贝和共享内存流水线的优化方案。现有实现仅支持标准掩码，无法处理业务中的非标准稀疏掩码，且全局内存访问不连续导致效率低下。我们通过构造连续对齐的掩码内存布局满足异步拷贝要求，并在有限共享内存中精细设计数据流策略，即前向计算缓存掩码块，反向计算实现掩码与中间变量分时复用，从而将掩码加载延迟隐藏于计算流水线中。
% 其次，我们基于Triton框架实现了Group Layer Normalization（GLN）和动态掩码构建的融合算子。原生PyTorch实现采用Gather-Compute-Scatter模式处理GLN，并通过碎片化操作构造动态掩码，导致频繁的kernel启动开销和低效的内存访问模式。基于Triton框架，我们通过算子融合策略消除中间结果的反复读写，利用向量化内存访问提升带宽利用率，并设计分组并行计算模式以充分发挥GPU并行性。实验表明，融合算子相比原生实现显著降低了kernel启动开销和内存访问延迟，有效消除了训练流水线中的计算瓶颈。
We conduct comprehensive optimizations to the offline training framework:

\textbf{CPU-GPU Pipeline Optimization. }In the domain of Large Language Models (LLMs), text undergoes tokenization with all features residing in GPU memory, enabling efficient end-to-end training. 
In contrast, recommendation systems face substantial feature engineering overhead, with numerous features stored in CPU memory. 
This architectural discrepancy creates blocking issues between GPU-resident models and CPU-resident features. 
The feature processing and model fusion pipelines involve extensive Host-Device data synchronization and serial dependencies, introducing multiple synchronization points where the Host must wait for Device-side operations to complete before proceeding. 
These CPU-GPU pipeline stalls severely degrade overall training performance.

To address this challenge, we systematically profiled the training framework using CUDA Profiler to identify CPU/GPU blocking bottlenecks. Our optimization strategy encompasses two key aspects: (1) eliminating synchronization points to enable CPU-GPU execution overlap, and (2) optimizing Device-side memory operations to reduce Device-to-Device (D2D) copy overhead. 
To implement the latter, we optimized frequent tensor indexing and assignment operations by consolidating multiple D2D copy operations into single atomic operations. 
Experimental results demonstrate that these systemic refinements improve training throughput by 20\%.

\textbf{Custom Kernel Development.}
First, we present an enhanced FlashAttention-2~\cite{dao2023flashattention} kernel designed to overcome the limitations of dynamic masking. 
While existing implementations are primarily optimized for standard causal masks, they often suffer from significant memory inefficiencies due to discontinuous access patterns when handling dynamic masks. 
To resolve this, we introduce contiguous and aligned mask layouts that facilitate high-throughput asynchronous copying. 
We also employ a fine-grained shared memory management strategy: we utilize available capacity to cache mask tiles during the forward pass, and implement temporal multiplexing of masks and intermediate tensors during the backward pass. 
This orchestration effectively overlaps mask-loading latency with the computation pipeline.

Furthermore, we implement fused operators for Group Layer Normalization (GLN) and dynamic mask construction using the Triton~\cite{tillet2019triton}. 
Native PyTorch implementations employ a Gather-Compute-Scatter pattern for GLN and fragmented operations for dynamic mask generation, resulting in excessive kernel launch overhead and inefficient memory access patterns. Leveraging Triton, we eliminate redundant intermediate memory transactions through operator fusion, enhance bandwidth utilization via vectorized memory access, and design group-wise parallel computation schemes to fully exploit GPU parallelism. 

Through these kernel-level optimizations, we achieved an additional 57\% throughput improvement.

\subsubsection{Inference Optimization}
% 为了提升MTFM模型在大规模工业场景下的推理效率，我们从算子级、模型级到系统部署级实施了一系列优化策略：
% 基于Ampere架构的结构化稀疏（Structured Sparsity）： 利用NVIDIA Ampere架构的稀疏张量核心（Sparse Tensor Cores），我们在HSTU组件的线性投影层（Q/K/V及输出层）实施了2:4结构化剪枝。该方法在保持模型精度的同时，实现了50%的显存压缩，并利用双倍理论峰值算力显著减少了矩阵乘法的计算延迟。
% 细粒度注意力计算剪枝（Fine-grained Attention Pruning）： 针对动态掩码中的不规则稀疏性，我们设计了计算跳过机制。该机制不仅动态剔除序列填充（Padding）产生的无效计算，还结合业务先验知识（如屏蔽用户特征对目标特征的无效注意力权重），相比标准因果掩码显著降低了注意力机制的计算开销。
% 场景化子图部署与系统级加速（Scenario-aware Deployment & Acceleration）： 由于MTFM是基于多个场景统一训练的，但是会被部署在各个独立子场景上，我们将完整计算图拆解为定制化子图以提升资源利用率，消除了本场景外的针对其他场景特征的额外计算。此外，结合BF16半精度推理与M-Falcon智能批处理算法，在平衡精度的前提下进一步最大化了推理吞吐量。

To enhance the inference efficiency of the MTFM model in large-scale industrial scenarios, we propose a comprehensive optimization framework spanning the sparsity, kernel, and system levels:

\textbf{Sparse Matrix Multiplication.} 
Leveraging the Sparse Tensor Cores within the NVIDIA Ampere architecture, we implement 2:4 structured pruning on the linear projection layers (UVQK and Output) of the HSTU module. 
This technique achieves a 50\% reduction in memory footprint and exploits double the theoretical peak throughput for matrix multiplication, significantly reducing latency without sacrificing predictive performance.
This optimization yields approximately 10\% throughput improvement and reduces inference latency by approximately 0.2ms.

\textbf{Attention Computation Optimization.} 
To address the irregular sparsity inherent in dynamic masks, we propose a fine-grained computation skipping strategy.
This approach dynamically prunes redundant operations stemming from sequence padding and leverages domain-specific constraints to bypass invalid computation (e.g., zero-weight attention from user to target features).
This optimization yields approximately 5\% throughput improvement.

% \textbf{Group Layer Normalization Optimization. }While the training phase focuses on operator fusion and memory bandwidth optimization, the inference phase incorporates a mask-based dynamic pruning mechanism. By exploiting token-group affiliations to construct dynamic masks, we implement fine-grained early termination for invalid group computations at the kernel initialization stage. This strategy effectively reduces the algorithmic time complexity from 
% O(G×N) to O(N) (where G denotes the number of groups). Consequently, this approach eliminates redundant operations at the algorithmic level while maintaining high parallelism, leading to a substantial reduction in inference latency.

\textbf{Scenario-aware Deployment and Other Optimizations.}
Since MTFM is trained across multiple scenarios but deployed on independent sub-scenarios, we decompose the global computational graph into scenario-specific subgraphs.
This modularization eliminates redundant operations associated with unnecessary scenario-specific features, thereby optimizing resource utilization.
Furthermore, we leverage BF16 half-precision inference and the M-Falcon~\cite{zhai2024actions} algorithm to accelerate inference and maximize system throughput.
\section{Experiments}
% 在本节中，我们会在真实世界的生产数据集上进行大量的实验来验证我们提出的MTFM模型的有效性并回答以下几个研究问题：

In this section, we conduct extensive experiments on our real- world production datasets to validate the effectiveness of our proposed framework MTFM and answer the following research questions:

% RQ1：MTFM模型在多个异构多场景上的效果和其他的基线对比效果如何？
\textbf{RQ1}: How does the MTFM model perform across multiple heterogeneous multi-scenario tasks compared to other baseline models?

% RQ2:  MTFM中混合注意力架构对模型训练推理性能的影响有多大？
\textbf{RQ2}: How does the hybrid attention architecture in MTFM affect model training and inference performance?

% RQ3: MTFM模型在扩展性上表现如何？包括模型大小扩展性以及数据扩展性。
\textbf{RQ3}: How does the MTFM model perform in terms of scalability, including model computational complexity scalability and data scalability?

% RQ4:MTFM可以体现跨场景token交互效果吗？
\textbf{RQ4}:Can MTFM provide interpretable insights into cross-scenario token interactions?

% RQ4: MTFM在生产环境中对各异构场景推荐的在线效果如何？
\textbf{RQ5}: What is the online recommendation performance of MTFM for various heterogeneous scenarios in production environments?

\subsection{Experiments Setup}
\subsubsection{Datasets}
\begin{table}[t]
\centering
\caption{Dataset statistics across our three scenarios.}
\label{tab:data}
\small 
\setlength{\tabcolsep}{3pt} 
\begin{tabularx}{\columnwidth}{lXXXXX}
\toprule
\textbf{Scenario} & \textbf{\#Users} & \textbf{\#Item} & \textbf{\#Exposure} & \textbf{\#Click} & \textbf{\#Purchase} \\ 
\midrule
HP  & 240M & 4.23M & 18.53B & 1.08B   & 176.77M \\
PHF & 151M & 8.07M & 15.29B & 359.14M & 104.73M \\
SQS & 44M  & 0.98M & 2.24B  & 85.34M  & 9.92M   \\
\bottomrule
\end{tabularx}
\vspace{-10px}
\end{table}
To accommodate the heterogeneous multi-scenario business data proposed in this work, we construct our offline experimental dataset from a real-world industrial food delivery recommendation system (Meituan) using user behavior logs. The dataset comprises three primary recommendation scenarios: \textbf{Homepage (HP)} Restaurant Recommendation, \textbf{Shenqiangshou(SQS)} Coupon-Package Recommendation , and \textbf{Pinhaofan(PHF)} Food Recommendation, with their respective data statistics summarized in Table \ref{tab:data}. Specifically, HP serves as the highest-exposure scenario where restaurant cards are recommended to users for order placement, while the SQS and PHF scenarios are food-centric with distinct business objectives (detailed in \ref{sub:metrics}). Over the past years, these scenarios have been continuously optimized both in business operations and recommendation architecture, making them the focal points for model innovation and industrial impact within Meituan’s ecosystem.
% Although these scenarios differ in their business objectives and development stages, the dining behaviors across all scenarios consistently reflect users' food preferences, which forms the foundation for our multi-scenario joint modeling approach.
%为了适配本文所提到的异构多场景业务数据，我们从美团的真实工业级外卖推荐系统中选取了从2025年12月23日-2025年1月12号超过两周的用户行为日志来构建我们的离线实验数据集。数据集包含了外卖推荐的三个最主要场景，首页商家推荐场景，神抢手券包推荐场景（SQS），拼好饭菜品拼单推荐场景（PHF），各自场景的相关数据统计分布信息如表【】所示。其中商家推荐场景是曝光流量最大的场景，给用户推荐外卖商家的相关卡片以供用户点击进入商家进行点单。而SQS和PHF场景则是和首页异构的，均是以菜品为载体。各个场景由于业务发展的阶段以及业务目标不尽相同，所以建模目标也有所差异，这部分目标说明我们在4.4.2中详细说明。但是他们用户在这些异构场景的点单行为都反映了他们的餐饮兴趣，这是我们多场景联合建模的基础。
\subsubsection{Metrics \& Baselines}
\label{sub:metrics}
We evaluate model performance primarily using AUC (Area Under the Curve) and GAUC (Grouped AUC) to assess prediction quality across diverse scenarios. 
For HP and PHF, we focus on CTR and CTCVR to measure exposure and conversion effectiveness. 
In the SQS, we extend our evaluation to include package redemption efficiency via two specialized metrics: IMD (Immediate Redemption within 30 minutes) and WRITE (Redemption within 24 hours).

To ensure rigorous comparison, we categorize baseline models according to their architecture types:
1) General recommenders including \textbf{DCNv2} \cite{DCNv2}, \textbf{MMoE} \cite{2018mmoe}, and \textbf{RankMixer} \cite{zhu2025rankmixer}, which are widely adopted for feature interaction and multi-task learning.
2) Generative Ranking Models, such as \textbf{MTGR} \cite{han2025mtgr} and \textbf{OneTrans} \cite{zhang2025onetrans}, representing the latest advances in generative sequence modeling.
3) Multi-Scenario Recommendation Models, such as \textbf{STAR} \cite{sheng2021star} and \textbf{PEPNet} \cite{chang2023pepnet}, which integrate multiple scenarios into a unified framework, allowing adaptive scenario modeling.
Notably, models in (1) and (2) are trained on a single-scenario dataset, whereas (3) and (4) leverage a multi-scenario dataset for training.

\subsubsection{Hyper-parameter Settings.}
For offline experiments, we use the Adam optimizer (learning rate $3 \times 10^{-4}$) with hyperparameters set to $d_{model} = 768, B = 4, K = 3, H=3$, and $G=1$.

% \newcolumntype{C}{>{\centering\arraybackslash}X}
\newcolumntype{Y}{>{\centering\arraybackslash}X}
\begin{table*}[!]
\renewcommand\arraystretch{0.72}            
\small
\centering
\caption{Performance comparison of different methods in terms of their respective task metrics on three scenarios. The best and second-best results are highlighted in boldface and underlined, respectively.}
  \vspace{-5px}
\label{tab::performanceHP_PHF}
% \begin{tabularx}{\textwidth}{@{}llcccccccc@{}}
\renewcommand\tabularxcolumn[1]{m{#1}} % 垂直居中
\begin{tabularx}{\textwidth}{llYYYYYYYY}
\toprule
\multicolumn{1}{c}{} & \multicolumn{1}{c}{} & \multicolumn{4}{c}{\textbf{Homepage Recommendation (HP)}} & \multicolumn{4}{c}{\textbf{Food Recommendation (PHF)}} \\ \cmidrule(l){3-6} \cmidrule(lr){7-10}
\multicolumn{2}{c}{\multirow{-3}{*}{\textbf{Method}}} & \multicolumn{2}{c}{CTR} & \multicolumn{2}{c}{CTCVR} & \multicolumn{2}{c}{CTR} & \multicolumn{2}{c}{CTCVR} \\ 
\cmidrule(l){3-4} \cmidrule(l){5-6} \cmidrule(l){7-8} \cmidrule(l){9-10} 
 \multicolumn{1}{c}{} & \multicolumn{1}{c}{}  & AUC & GAUC & AUC & GAUC & AUC & GAUC & AUC & GAUC \\ \midrule
 & DCNv2 & 0.7664 & 0.6853 & 0.8780 & 0.6451 & 0.7683 & 0.7236 & 0.8586 & 0.7555 \\
 & MMoE & 0.7664 & 0.6857 & 0.8782 & 0.6454 & 0.7718 & 0.7282 & 0.8640 & 0.7597 \\
\multirow{-3}{*}{General Recommenders} & Rankmixer & 0.7665 & 0.6860 & 0.8789 & 0.6464 & 0.7711 & 0.7270 & 0.8628 & 0.7590 \\ \midrule
 & OneTrans & 0.7672 & 0.6944 & 0.8774 & \ul{ 0.6497} & 0.7832 & 0.7373 & 0.8827 & 0.7735 \\
\multirow{-2}{*}{Generative Ranking Method} & MTGR & \ul{0.7679} & \ul{ 0.6951} & 0.8776 & 0.6491 & \ul{ 0.7883} & \ul{ 0.7398} & 0.8879 & \ul{ 0.7771} \\ \midrule
 & STAR & 0.7669 & 0.6882 & 0.8780 & 0.6482 & 0.7821 & 0.7298 & 0.8688 & 0.7660 \\
\multirow{-2}{*}{Multi-Scenario Method} & PEPNet & 0.7672 & 0.6895 & \ul{ 0.8790} & 0.6489 & 0.7866 & 0.7328 & 0.8721 & 0.7693 \\
\midrule
\textbf{Our Proposed Method} & \textbf{MTFM} & \textbf{0.7689} & \textbf{0.6954} & \textbf{0.8806} & \textbf{0.6507} & \textbf{0.7940} & \textbf{0.7474} & \textbf{0.8892} & \textbf{0.7824} \\ 
\bottomrule
\end{tabularx}
\end{table*}
\newcolumntype{Y}{>{\centering\arraybackslash}X}
\begin{table*}[!]
\renewcommand\arraystretch{0.72}
\small
\centering
% \begin{tabular}{@{}llcccccccc@{}}
\renewcommand\tabularxcolumn[1]{m{#1}} % 垂直居中
\begin{tabularx}{\textwidth}{llYYYYYYYY}
\toprule
\multicolumn{1}{c}{} & \multicolumn{1}{c}{} & \multicolumn{8}{c}{\textbf{Coupon-Package Recommendation (SQS)}} \\ 
\cmidrule(l){3-10}
\multicolumn{2}{c}{\multirow{-3}{*}{\textbf{Method}}} & \multicolumn{2}{c}{CTR} & \multicolumn{2}{c}{CTCVR} & \multicolumn{2}{c}{IMD} & \multicolumn{2}{c}{WRITE} \\ 
\cmidrule(l){3-4} \cmidrule(l){5-6} \cmidrule(l){7-8} \cmidrule(l){9-10} 
 \multicolumn{1}{c}{} & \multicolumn{1}{c}{}  & AUC & GAUC & AUC & GAUC & AUC & GAUC & AUC & GAUC \\ \midrule
 & DCNv2 & 0.8290 & 0.7789 & 0.9057 & 0.8227 & 0.9074 & 0.8254 & 0.9072 & 0.8222 \\
 & MMoE & 0.8449 & 0.7842 & 0.9073 & 0.8243 & 0.9082 & 0.8252 & 0.9056 & 0.8230 \\
\multirow{-3}{*}{General Recommenders} & Rankmixer & 0.8472 & 0.7881 & 0.9094 & 0.8281 & 0.9105 & 0.8268 & \textbf{0.9080} & \ul{ 0.8279} \\ \midrule
 & OneTrans & 0.8454 & 0.7994 & 0.9091 & 0.8271 & 0.9089 & 0.8279 & 0.9055 & 0.8248 \\
\multirow{-2}{*}{Generative Ranking Method} & MTGR & 0.8454 & \ul{ 0.7997} & \ul{ 0.9097} & \ul{ 0.8282} & \ul{ 0.9095} & \ul{ 0.8291} & 0.9059 & 0.8258 \\ \midrule
 & STAR & 0.8470 & 0.7863 & 0.9079 & 0.8245 & 0.9081 & 0.8257 & 0.9055 & 0.8233 \\
\multirow{-2}{*}{Multi-Scenario Method} & PEPNet & \ul{ 0.8511} & 0.7892 & 0.9081 & 0.8251 & 0.9089 & 0.8268 & 0.9066 & 0.8239 \\ \midrule
\textbf{Our Proposed Method} & \textbf{MTFM} & \textbf{0.8624} & \textbf{0.8027} & \textbf{0.9119} & \textbf{0.8301} & \textbf{0.9117} & \textbf{0.8319} & \ul{ 0.9079} & \textbf{0.8288} \\ \bottomrule
\end{tabularx}
\end{table*}

\subsection{Overall Performance Comparison (RQ1)}

Table \ref{tab::performanceHP_PHF} presents the main experimental results for three representative scenarios: HP, PHF, and SQS. 
Several key insights can be drawn. 
First, RankMixer delivers competitive results among general recommenders by effectively supporting model scaling; it marginally exceeds other baselines on HP while achieving substantial gains on SQS, outperforming MMoE by 0.38pp in CTCVR GAUC and 0.49pp in WRITE GAUC. However, in the PHF setting, its performance remains superior to DCNv2 but trails behind MMoE. 
Second, multi-scenario methods such as STAR and PEPNet generally outperform general baselines but fall short of generative ranking models like MTGR and OneTrans. Notably, these methods exhibit a pronounced "see-saw effect"; for instance, PEPNet surpasses RankMixer in HP and PHF but underperforms in the SQS scenario. 
In contrast, generative models consistently demonstrate superior robustness, with MTGR slightly edging out OneTrans across most metrics.
Ultimately, our proposed MTFM model achieves state-of-the-art results across nearly all scenarios and tasks. 
The consistent dominance of MTFM underscores its capability to not only mitigate the see-saw effect but, more importantly, to capitalize on the scaling law by efficiently leveraging extensive multi-scenario data to further enhance model's performance.

\subsection{Efficiency Analysis of HTA (RQ2)}

% \begin{table}[t]
% \centering
% \caption{Performance comparison of hybrid architecture configurations with varying target-to-full attention ratios. "w/o opt" denotes without system-level engineering optimizations (Section~\ref{sec:optimization}). Throughput is measured as samples per second on a single NVIDIA A100 GPU.}
% \label{tab:hybrid_config}
% \scalebox{0.88}{
% \begin{tabular}{lcccc}
% \toprule
% \textbf{Configuration} & \textbf{AUC} & \textbf{GAUC} & \textbf{Throughput} & \textbf{Memory} \\
% \textbf{(Target:Full)} & & & \textbf{(samples/s)} & \textbf{(GB)} \\
% \midrule
% 0:16 (w/o opt) & 0.7514 & 0.6818 & 220 & 66.32 \\
% 0:16 (Full-Only) & 0.7514 & 0.6818 & 390 & 66.97 \\
% \midrule
% (1:1)*8 & 0.7508 & 0.6820 & 497 & 38.00 \\
% (3:1)*4 & 0.7506 & 0.6821 & 547 & 34.08 \\
% (3:1)*4 (bs*2) & 0.7506 & 0.6822 & 780 & 67.49 \\
% (3:1)*4 (bs*2 w/o GQA) & 0.7506 & 0.6822 & 660 & 70.16\\
% (5:1)*3 & 0.7504 & 0.6811 & 535 & 37.52 \\
% \midrule
% 16:0 (Target-Only) & 0.7497 & 0.6806 & 575 & 32.64 \\
% \bottomrule
% \end{tabular}}
% \end{table}

\begin{table}[t]
\centering
\caption{Performance comparison of hybrid architecture configurations with varying target-to-full attention ratios. 
% "w/o opt" denotes without system-level engineering optimizations (Section~\ref{sec:optimization}). 
Throughput is measured as samples per second on a single NVIDIA A100 GPU.}
\label{tab:hybrid_config}
\scalebox{0.82}{
\begin{tabular}{lccccc}
\toprule
\textbf{Configuration} & \textbf{Batch} & \textbf{AUC} & \textbf{GAUC} & \textbf{Throughput} & \textbf{Memory} \\
\textbf{(Target:Full)} & \textbf{Size} & & & \textbf{(samples/s)} & \textbf{(GB)} \\
\midrule
\multicolumn{6}{l}{\textit{Full Attention Only}} \\
% 0:16 (w/o opt) & 1$\times$ & 0.7514 & 0.6818 & 220 & 66.32 \\
(0:16)*1  & 1$\times$ & 0.7514 & 0.6818 & 390 & 66.97 \\
\midrule
\multicolumn{6}{l}{\textit{Target Attention Only}} \\
(16:0)*1  & 1$\times$ & 0.7497 & 0.6806 & 575 & 32.64 \\

\midrule
\multicolumn{6}{l}{\textit{Hybrid Configurations}} \\
(1:1)*8 & 1$\times$ & 0.7508 & 0.6820 & 497 & 38.00 \\
(3:1)*4 & 1$\times$ & 0.7506 & 0.6821 & 547 & 34.08 \\
(5:1)*3 & 1$\times$ & 0.7504 & 0.6811 & 535 & 37.52 \\
(3:1)*4 (w/o GQA) & 2$\times$ & 0.7506 & 0.6822 & 660 & 70.16\\

\textbf{(3:1)*4 (MTFM) }& 2$\times$ & 0.7506 & 0.6822 & \textbf{780} & 67.49 \\

\bottomrule
\end{tabular}}
\end{table}
% 在本节，我们对混合架构的不同实验设置进行了探索。我们选取了美团外卖场景的7天数据样本进行了实验。
% 如表所示，对于baseline模型，我们的工程优化使得整体训练速度提升了77%。
% 对于混合架构实验，我们对比了target:full的不同比例的模型性能。如表所示，1:1和3:1的混合架构性能基本无损（0.1pp以内），在提速的同时显著降低了显存占用。而5:1存在轻度的性能下跌，并且纯target attention结构下跌更为明显。此外，GQA可以进一步加速并降低显存。在实际模型架构选择上，我们选择了3:1的GQA方案，并且扩大batch_size实现进一步加速
In this section, we conducted extensive experiments to explore different configurations of the hybrid target attention. We performed experiments using another 7-day sample for training. 
% 我们记(P:K)*N表示不同的网络架构配置，其N代表了总的block数，每一个block中，P代表了Full Attention层的数量，K代表了target Attention层的数量。我们还额外比较了一种特殊情况，以及target attention only，此时模型退化成onerecV2种lazy decoder only。
We denote the various network architecture configurations as $(K:P) \times B$, where $B$ represents the total number of blocks. Within each block, $K$ and $P$ signify the number of Target Attention layers and Full Attention layers, respectively.
We further evaluate a special case, Target Attention Only, where the model degenerates into the lazy decoder as seen in OneRecV2\cite{zhou2025onerecv2technicalreport}.
% As shown in Table~\ref{tab:hybrid_config}, our training optimizations in Section~\ref{sec:optimization} achieved a 77\% improvement in overall training speed for the baseline model.

% We compared model performance across different ratios of target-to-full attention layers. As shown in Table~\ref{tab:hybrid_config}, the 1:1 and 3:1 hybrid architectures exhibited negligible performance degradation (within 0.1 percentage points), while significantly reducing memory usage and accelerating training. The 5:1 configuration showed mild performance degradation, with a more pronounced decline observed in the pure target attention architecture. Furthermore, experimental results show that GQA  can further enhance computational efficiency.

We evaluate the impact of varying the ratio between target and full attention layers (Table \ref{tab:hybrid_config}).
The 1:1 and 3:1 configurations maintain or even slightly improve performance relative to full attention while delivering significant gains in both memory efficiency and training throughput.
When the sparsity ratio is further increased to 5:1, performance begins to decline by 0.07pp, with the degradation widening to 0.12pp under a pure target attention setup. 
Additionally, the integration of GQA provides further gains in throughput alongside additional reductions in memory consumption. 
Finally, our HTA achieves a 2x speedup in training throughput compared to the full attention mechanism.

\subsection{Scalability (RQ3)}
In this section, we systematically evaluate the scalability of MTFM for multi-scenario recommendation (MSR) by examining its performance across varying model scales and training data volumes. 
% \input{tabs/tab.comp_complexity}

% 我们设置了不同的模型尺寸(using MMoE as the baseline)来验证MTFM的Scalability。这里我们采用了CTCVR GAUC作为metric，模型尺寸横跨10x-70x，并同时在SQS、HP、PHF三个场景进行对比。As shown in Figure \ref{fig:mtfm_scalability}(a), MTFM在所有场景都展现了良好的scaling law，并且具有一致的scaling law斜率，这些都表明了MTFM在应对MRS时对于model size具有较好的Scalability。
\subsubsection{Scalability with Model Size in MSR}
We evaluated its performance across various model scales (ranging from 10x to 70x), using MMoE as the competitive baseline. 
We employ CTCVR GAUC as the primary metric across three distinct scenarios: SQS, HP, and PHF. 
As illustrated in Figure \ref{fig:mtfm_scalability}(a), MTFM exhibits robust scaling laws with consistent slopes across all scenarios, underscoring its superior scalability regarding model capacity in MSR tasks.
% As shown in Figure \ref{fig:mtfm_scalability}(a), we examine the scaling law of CTCVR GAUC gain with respect to relative inference FLOPs, using DLRM as the baseline (DLRM = 1×). Across all three scenarios (SQS, HP, and PHF), a strong positive correlation is observed between increased computational resources and offline model performance. Notably, the fitted dotted lines and high $R^2$ values (>0.93) in each scenario reflect stable and consistent scaling trends. 
% These results demonstrate that MTFM can efficiently leverage additional inference capacity to achieve substantial improvements in sequence modeling and feature interaction accuracy.

% 我们进一步分析了数据规模对于MTFM的影响，我们选择了三种不同的尺寸模型，观察他们持续训练时对于CTCVR GAUC的表现变化。
\subsubsection{Scalability with Training Data Volumes in MRS}
\begin{figure}[t]
    \centering
    % 横向并排，宽度各为0.48\textwidth
    \begin{subfigure}[b]{0.48\textwidth}
        \centering
        \includegraphics[width=\textwidth]{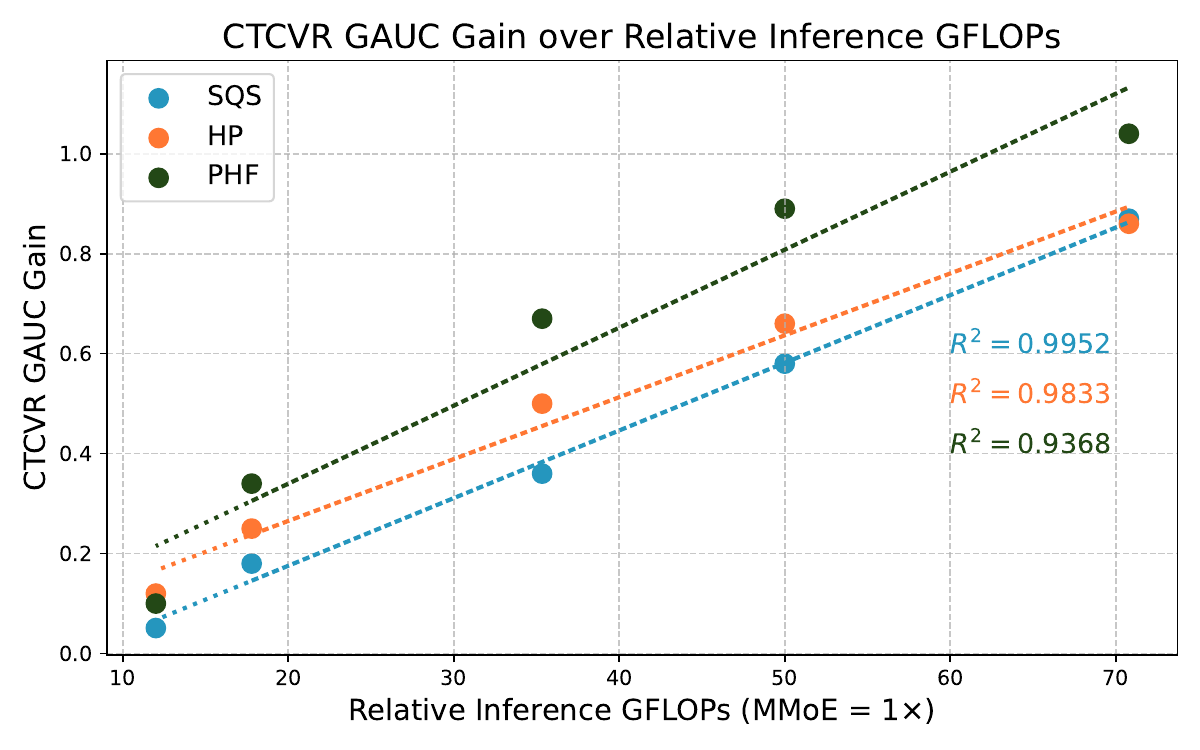}
        \caption{CTCVR GAUC scaling with inference GFLOPs across 3 scenarios.}
        \label{fig:scatter_fit}
    \end{subfigure}
    \hspace{0.02\textwidth}
    \begin{subfigure}[b]{0.48\textwidth}
        \centering
        \includegraphics[width=\textwidth]{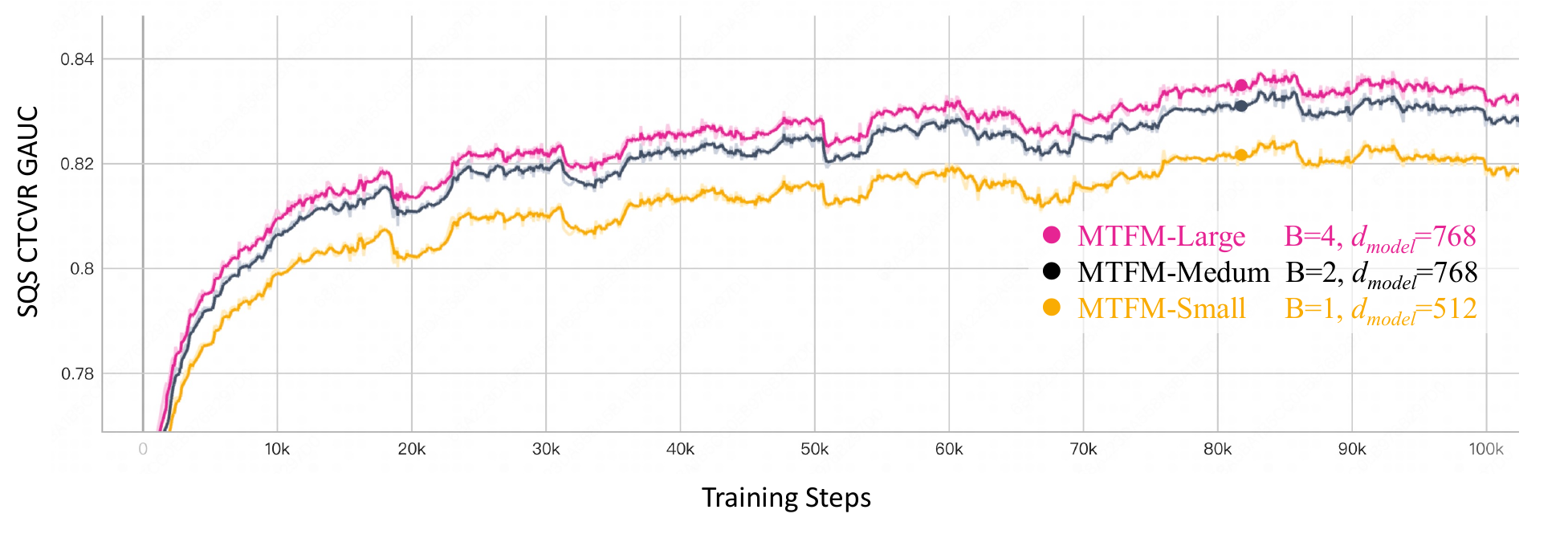}
        \caption{GAUC scaling with training data for different model sizes.}
        \label{fig:train_auc}
    \end{subfigure}
    \caption{
        Scalability analysis of MTFM.
        % \textbf{(a)} CTCVR GAUC gain as a function of relative inference GFLOPs in SQS, HP, and PHF scenarios.
        % \textbf{(b)} GAUC scaling with training data volume for small, medium, and large MTFM variants.
    }
    \label{fig:mtfm_scalability}
\end{figure}

As depicted in Figure \ref{fig:mtfm_scalability}(b), we investigate the impact of the data scale on MTFM by selecting models of three different sizes and observing their performance fluctuations in SQS's CTCVR GAUC throughout the continual training phase. 
Empirical results reveal that across all model scales, performance improves consistently as the volume of training tokens increases. 
In experiments with larger data scales, we further observe that the performance gap between models of different sizes widens gradually as the number of training tokens increases.
% 在更大规模数据的实验中，我们还观察到不同尺寸模型的性能差异会伴随训练token增多而缓慢增加。This robust scaling trend underscores MTFM’s 关于训练数据量上的scalability
This robust scaling trend underscores MTFM’s data scalability and demonstrates its ability to consistently convert increased training volumes into performance improvements.

% The large model maintains a stable advantage throughout, reflecting its enhanced ability to capture complex feature interactions and sequence dependencies as data availability grows. 

% Overall, these comprehensive experiments confirm that MTFM prioritizes scaling efficiency: by increasing computational resources or model size, MTFM realizes greater performance gains per parameter and per training token, leading to more rapid improvements than conventional baselines.

\subsection{Interpretability Analysis (RQ4)}
\begin{figure}[htbp]
    \centering
    \includegraphics[width=\columnwidth]{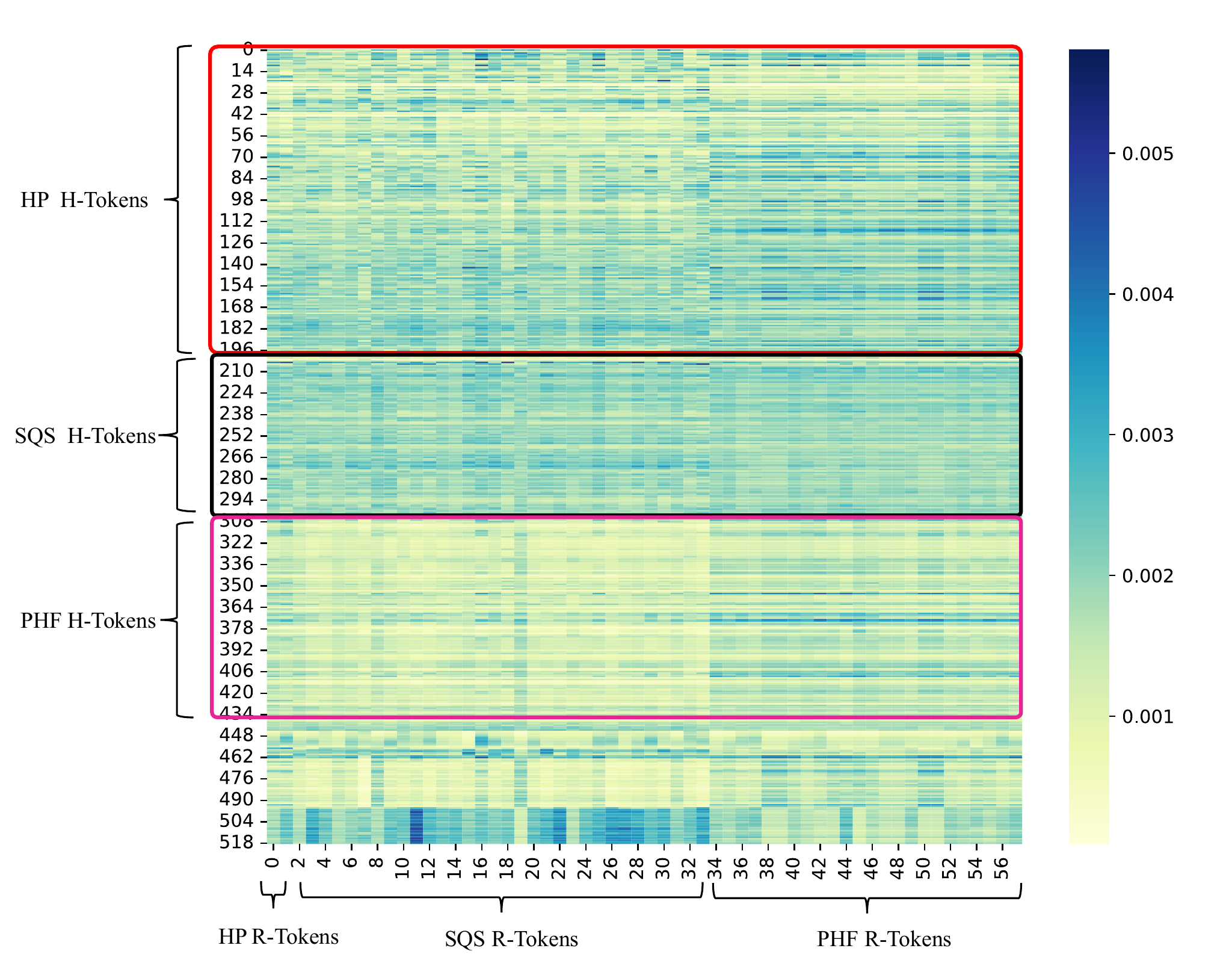}
    \caption{
        Attention heatmap visualization for MTFM. 
        The horizontal/vertical axis represents T/H-tokens respectively.
    }
    \label{fig:att_heatmap}
    \vspace{-10px}
\end{figure}
% To understand how MTFM models scenario-specific semantic interactions, we visualize the attention heatmap of a training sample, focusing on the horizontal axis target token regions.

% Figure \ref{fig:att_heatmap} shows that HP and SQS target tokens exhibit higher attention to their respective historical than  PHF historical tokens. In contrast, PHF target tokens more strongly attend to PHF history, with both HP and SQS histories also playing supplementary roles for PHF recommendations.Overall, the heatmap reveals clear block-wise attention patterns: each target scenario concentrates attention on the most relevant item features, aligning with business intuition and demonstrating that MTFM’s attention mechanism is interpretable and scenario-aware.

To understand how MTFM models scenario-specific characteristics and captures differentiated cross-scenario knowledge, we visualize the attention heatmaps for representative samples in Figure \ref{fig:att_heatmap}. 
The horizontal axis denotes T-Tokens across scenarios, while the vertical axis represents H-Tokens. 
Our observations include: 1) T-Tokens in all scenarios effectively aggregate information from H-Tokens across the entire multi-scenario space. 2) Distinct learning patterns emerge across scenarios; for instance, T-Tokens in HP and SQS exhibit higher attention weights toward their own H-Tokens compared to those of PHF, whereas PHF T-Tokens are more heavily indexed on their corresponding H-Tokens.
Overall, the heatmaps reveal distinct block-wise attention patterns, demonstrating that MTFM is not only capable of learning from multi-scenario data but also possesses robust scenario-awareness.

\subsection{Online Experiments (RQ5)}
To validate MTFM's real-world impact, we conducted A/B tests across two major online scenarios. 
The experimental traffic involved tens of millions of daily exposures, ensuring robust results. 
The baseline model was the current SOTA model, which has been continuously optimized and deployed for several years.

\begin{table}[h]
\centering
\small
\caption{Online Performance of MTFM.}
\label{tab:online_exp}
\vspace{-2mm} 
\begin{tabular}{lccccc}
\toprule
Scenario & CTR & UV\_CTCVR & ORDERS & LATENCY\\ 
\midrule
SQS   & +1.89\% & +2.46\% & +2.98\% & -5ms \\
PHF   & +1.53\% & +1.03\% & +1.45\% & -6ms \\
\bottomrule
\end{tabular}
% \vspace{-5px}
\end{table}

As shown in Table \ref{tab:online_exp}, MTFM delivers consistent performance improvements across all key business metrics.
% MTFM reduces online latency due to reusing of user sequence computations in incoming requests.
% Notably, in the PHF scenario, an increase of just 0.1\% in order volume is considered a significant business improvement. 
The observed online uplift in orders with MTFM is equivalent to the cumulative gains typically achieved over 2–3 rounds of model iteration in this business domain. 
In addition, we evaluate the latency of MTFM, where the results indicate that our system optimizations lead to better performance.
% All metrics exhibited stable improvements during testing. 
These results clearly demonstrate that MTFM could deliver direct business value in real-world recommendation systems.

\section{Conclusion}
In this paper, we propose MTFM, a recommendation foundation model that leverages heterogeneous tokenization to achieve a truly alignment-free architecture across diverse domains. Through the co-design of algorithms, training, and deployment systems, MTFM simultaneously embodies the three quintessential properties: Scalability, Extensibility, and Efficiency. Extensive offline evaluations demonstrate significant performance gains, establishing a new scalable paradigm that successfully transcends the limitations of single-scenario silos. Finally, online A/B tests confirm MTFM's real-world effectiveness, delivering substantial improvements in CTR and conversion volumes in Meituan.
% \begin{acks}
% \end{acks}

\bibliographystyle{ACM-Reference-Format}
\balance
\bibliography{ref}


\begin{thebibliography}{37}

%%% the Web address.

\ifx \showCODEN    \undefined \def \showCODEN     #1{\unskip}     \fi
\ifx \showISBNx    \undefined \def \showISBNx     #1{\unskip}     \fi
\ifx \showISBNxiii \undefined \def \showISBNxiii  #1{\unskip}     \fi
\ifx \showISSN     \undefined \def \showISSN      #1{\unskip}     \fi
\ifx \showLCCN     \undefined \def \showLCCN      #1{\unskip}     \fi
\ifx \shownote     \undefined \def \shownote      #1{#1}          \fi
\ifx \showarticletitle \undefined \def \showarticletitle #1{#1}   \fi
\ifx \showURL      \undefined \def \showURL       {\relax}        \fi
% invisible to TeX
\providecommand\bibfield[2]{#2}
\providecommand\bibinfo[2]{#2}
\providecommand\natexlab[1]{#1}
\providecommand\showeprint[2][]{arXiv:#2}

\bibitem[Ainslie et~al\mbox{.}(2023)]%
        {ainslie2023gqa}
\bibfield{author}{\bibinfo{person}{Joshua Ainslie}, \bibinfo{person}{James Lee-Thorp}, \bibinfo{person}{Michiel de Jong}, \bibinfo{person}{Yury Zemlyanskiy}, \bibinfo{person}{Federico Lebron}, {and} \bibinfo{person}{Sumit Sanghai}.} \bibinfo{year}{2023}\natexlab{}.
\newblock \showarticletitle{GQA: Training Generalized Multi-Query Transformer Models from Multi-Head Checkpoints}. In \bibinfo{booktitle}{\emph{Proceedings of the 2023 Conference on Empirical Methods in Natural Language Processing}}. \bibinfo{pages}{4895--4901}.
\newblock


\bibitem[Bai et~al\mbox{.}(2025)]%
        {bai2025qwen3vltechnicalreport}
\bibfield{author}{\bibinfo{person}{Shuai Bai}, \bibinfo{person}{Yuxuan Cai}, \bibinfo{person}{Ruizhe Chen}, \bibinfo{person}{Keqin Chen}, \bibinfo{person}{Xionghui Chen}, \bibinfo{person}{Zesen Cheng}, \bibinfo{person}{Lianghao Deng}, \bibinfo{person}{Wei Ding}, \bibinfo{person}{Chang Gao}, {et~al\mbox{.}}} \bibinfo{year}{2025}\natexlab{}.
\newblock \bibinfo{title}{Qwen3-VL Technical Report}.
\newblock
\showeprint[arxiv]{2511.21631}~[cs.CV]
\urldef\tempurl%
\url{https://arxiv.org/abs/2511.21631}
\showURL{%
\tempurl}


\bibitem[Chang et~al\mbox{.}(2023)]%
        {chang2023pepnet}
\bibfield{author}{\bibinfo{person}{Jianxin Chang}, \bibinfo{person}{Chenbin Zhang}, \bibinfo{person}{Yiqun Hui}, \bibinfo{person}{Dewei Leng}, \bibinfo{person}{Yanan Niu}, \bibinfo{person}{Yang Song}, {and} \bibinfo{person}{Kun Gai}.} \bibinfo{year}{2023}\natexlab{}.
\newblock \showarticletitle{Pepnet: Parameter and embedding personalized network for infusing with personalized prior information}. In \bibinfo{booktitle}{\emph{Proceedings of the 29th ACM SIGKDD Conference on Knowledge Discovery and Data Mining}}. \bibinfo{pages}{3795--3804}.
\newblock


\bibitem[Dao(2023)]%
        {dao2023flashattention}
\bibfield{author}{\bibinfo{person}{Tri Dao}.} \bibinfo{year}{2023}\natexlab{}.
\newblock \showarticletitle{Flashattention-2: Faster attention with better parallelism and work partitioning}.
\newblock \bibinfo{journal}{\emph{arXiv preprint arXiv:2307.08691}} (\bibinfo{year}{2023}).
\newblock


\bibitem[Deng et~al\mbox{.}(2025)]%
        {deng2025onerec}
\bibfield{author}{\bibinfo{person}{Jiaxin Deng}, \bibinfo{person}{Shiyao Wang}, \bibinfo{person}{Kuo Cai}, \bibinfo{person}{Lejian Ren}, \bibinfo{person}{Qigen Hu}, \bibinfo{person}{Weifeng Ding}, \bibinfo{person}{Qiang Luo}, {and} \bibinfo{person}{Guorui Zhou}.} \bibinfo{year}{2025}\natexlab{}.
\newblock \showarticletitle{Onerec: Unifying retrieve and rank with generative recommender and iterative preference alignment}.
\newblock \bibinfo{journal}{\emph{arXiv preprint arXiv:2502.18965}} (\bibinfo{year}{2025}).
\newblock


\bibitem[Fedus et~al\mbox{.}(2022)]%
        {fedus2022switch}
\bibfield{author}{\bibinfo{person}{William Fedus}, \bibinfo{person}{Barret Zoph}, {and} \bibinfo{person}{Noam Shazeer}.} \bibinfo{year}{2022}\natexlab{}.
\newblock \showarticletitle{Switch transformers: Scaling to trillion parameter models with simple and efficient sparsity}.
\newblock \bibinfo{journal}{\emph{Journal of Machine Learning Research}} \bibinfo{volume}{23}, \bibinfo{number}{120} (\bibinfo{year}{2022}), \bibinfo{pages}{1--39}.
\newblock


\bibitem[Fu et~al\mbox{.}(2025)]%
        {fu2025vita}
\bibfield{author}{\bibinfo{person}{Chaoyou Fu}, \bibinfo{person}{Haojia Lin}, \bibinfo{person}{Xiong Wang}, \bibinfo{person}{Yi-Fan Zhang}, \bibinfo{person}{Yunhang Shen}, \bibinfo{person}{Xiaoyu Liu}, \bibinfo{person}{Haoyu Cao}, \bibinfo{person}{Zuwei Long}, \bibinfo{person}{Heting Gao}, \bibinfo{person}{Ke Li}, {et~al\mbox{.}}} \bibinfo{year}{2025}\natexlab{}.
\newblock \showarticletitle{Vita-1.5: Towards gpt-4o level real-time vision and speech interaction}.
\newblock \bibinfo{journal}{\emph{arXiv preprint arXiv:2501.01957}} (\bibinfo{year}{2025}).
\newblock


\bibitem[Han et~al\mbox{.}(2024)]%
        {han2024enhancing}
\bibfield{author}{\bibinfo{person}{Ruidong Han}, \bibinfo{person}{Qianzhong Li}, \bibinfo{person}{He Jiang}, \bibinfo{person}{Rui Li}, \bibinfo{person}{Yurou Zhao}, \bibinfo{person}{Xiang Li}, {and} \bibinfo{person}{Wei Lin}.} \bibinfo{year}{2024}\natexlab{}.
\newblock \showarticletitle{Enhancing CTR Prediction through Sequential Recommendation Pre-training: Introducing the SRP4CTR Framework}. In \bibinfo{booktitle}{\emph{Proceedings of the 33rd ACM International Conference on Information and Knowledge Management}}. \bibinfo{pages}{3777--3781}.
\newblock


\bibitem[Han et~al\mbox{.}(2025)]%
        {han2025mtgr}
\bibfield{author}{\bibinfo{person}{Ruidong Han}, \bibinfo{person}{Bin Yin}, \bibinfo{person}{Shangyu Chen}, \bibinfo{person}{He Jiang}, \bibinfo{person}{Fei Jiang}, \bibinfo{person}{Xiang Li}, \bibinfo{person}{Chi Ma}, \bibinfo{person}{Mincong Huang}, \bibinfo{person}{Xiaoguang Li}, \bibinfo{person}{Chunzhen Jing}, {et~al\mbox{.}}} \bibinfo{year}{2025}\natexlab{}.
\newblock \showarticletitle{Mtgr: Industrial-scale generative recommendation framework in meituan}. In \bibinfo{booktitle}{\emph{Proceedings of the 34th ACM International Conference on Information and Knowledge Management}}. \bibinfo{pages}{5731--5738}.
\newblock


\bibitem[Hoffmann et~al\mbox{.}(2022)]%
        {hoffmann2022training}
\bibfield{author}{\bibinfo{person}{Jordan Hoffmann}, \bibinfo{person}{Sebastian Borgeaud}, \bibinfo{person}{Arthur Mensch}, \bibinfo{person}{Elena Buchatskaya}, \bibinfo{person}{Trevor Cai}, \bibinfo{person}{Eliza Rutherford}, \bibinfo{person}{Diego de~Las Casas}, \bibinfo{person}{Lisa~Anne Hendricks}, \bibinfo{person}{Johannes Welbl}, \bibinfo{person}{Aidan Clark}, {et~al\mbox{.}}} \bibinfo{year}{2022}\natexlab{}.
\newblock \showarticletitle{Training compute-optimal large language models}.
\newblock \bibinfo{journal}{\emph{arXiv preprint arXiv:2203.15556}} (\bibinfo{year}{2022}).
\newblock


\bibitem[Kaplan et~al\mbox{.}(2020)]%
        {kaplan2020scaling}
\bibfield{author}{\bibinfo{person}{Jared Kaplan}, \bibinfo{person}{Sam McCandlish}, \bibinfo{person}{Tom Henighan}, \bibinfo{person}{Tom~B Brown}, \bibinfo{person}{Benjamin Chess}, \bibinfo{person}{Rewon Child}, \bibinfo{person}{Scott Gray}, \bibinfo{person}{Alec Radford}, \bibinfo{person}{Jeffrey Wu}, {and} \bibinfo{person}{Dario Amodei}.} \bibinfo{year}{2020}\natexlab{}.
\newblock \showarticletitle{Scaling laws for neural language models}.
\newblock \bibinfo{journal}{\emph{arXiv preprint arXiv:2001.08361}} (\bibinfo{year}{2020}).
\newblock


\bibitem[Li et~al\mbox{.}(2025)]%
        {li2025realizing}
\bibfield{author}{\bibinfo{person}{Dai Li}, \bibinfo{person}{Kevin Course}, \bibinfo{person}{Wei Li}, \bibinfo{person}{Hongwei Li}, \bibinfo{person}{Jie Hua}, \bibinfo{person}{Yiqi Chen}, \bibinfo{person}{Zhao Zhu}, \bibinfo{person}{Rui Jian}, \bibinfo{person}{Xuan Cao}, \bibinfo{person}{Bi Xue}, {et~al\mbox{.}}} \bibinfo{year}{2025}\natexlab{}.
\newblock \showarticletitle{Realizing Scaling Laws in Recommender Systems: A Foundation-Expert Paradigm for Hyperscale Model Deployment}.
\newblock \bibinfo{journal}{\emph{arXiv preprint arXiv:2508.02929}} (\bibinfo{year}{2025}).
\newblock


\bibitem[Liu et~al\mbox{.}(2026)]%
        {liu2026ministral3}
\bibfield{author}{\bibinfo{person}{Alexander~H. Liu}, \bibinfo{person}{Kartik Khandelwal}, \bibinfo{person}{Sandeep Subramanian}, \bibinfo{person}{Victor Jouault}, \bibinfo{person}{Abhinav Rastogi}, \bibinfo{person}{Adrien Sadé}, \bibinfo{person}{Alan Jeffares}, \bibinfo{person}{Albert Jiang}, \bibinfo{person}{Alexandre Cahill}, {et~al\mbox{.}}} \bibinfo{year}{2026}\natexlab{}.
\newblock \bibinfo{title}{Ministral 3}.
\newblock
\showeprint[arxiv]{2601.08584}~[cs.CL]
\urldef\tempurl%
\url{https://arxiv.org/abs/2601.08584}
\showURL{%
\tempurl}


\bibitem[Ma et~al\mbox{.}(2018a)]%
        {ma2018modeling}
\bibfield{author}{\bibinfo{person}{Jiaqi Ma}, \bibinfo{person}{Zhe Zhao}, \bibinfo{person}{Xinyang Yi}, \bibinfo{person}{Jilin Chen}, \bibinfo{person}{Lichan Hong}, {and} \bibinfo{person}{Ed~H Chi}.} \bibinfo{year}{2018}\natexlab{a}.
\newblock \showarticletitle{Modeling task relationships in multi-task learning with multi-gate mixture-of-experts}. In \bibinfo{booktitle}{\emph{Proceedings of the 24th ACM SIGKDD international conference on knowledge discovery \& data mining}}. \bibinfo{pages}{1930--1939}.
\newblock


\bibitem[Ma et~al\mbox{.}(2018b)]%
        {2018mmoe}
\bibfield{author}{\bibinfo{person}{Jiaqi Ma}, \bibinfo{person}{Zhe Zhao}, \bibinfo{person}{Xinyang Yi}, \bibinfo{person}{Jilin Chen}, \bibinfo{person}{Lichan Hong}, {and} \bibinfo{person}{Ed~H. Chi}.} \bibinfo{year}{2018}\natexlab{b}.
\newblock \showarticletitle{Modeling Task Relationships in Multi-task Learning with Multi-gate Mixture-of-Experts}. In \bibinfo{booktitle}{\emph{Proceedings of the 24th ACM SIGKDD International Conference on Knowledge Discovery \& Data Mining}}. \bibinfo{pages}{1930–1939}.
\newblock


\bibitem[{Qwen Team}(2025)]%
        {qwen3next}
\bibfield{author}{\bibinfo{person}{{Qwen Team}}.} \bibinfo{year}{2025}\natexlab{}.
\newblock \bibinfo{title}{Qwen3-Next: Towards Ultimate Training and Inference Efficiency}.
\newblock \bibinfo{howpublished}{\url{https://qwen3-next.com/}}.
\newblock
\newblock
\shownote{Accessed: 2026-01-20}.


\bibitem[Recommendation(2025)]%
        {recommendation2025external}
\bibfield{author}{\bibinfo{person}{Ads Recommendation}.} \bibinfo{year}{2025}\natexlab{}.
\newblock \showarticletitle{External Large Foundation Model: How to Efficiently Serve Trillions of Parameters for Online Ads Recommendation}.
\newblock \bibinfo{journal}{\emph{arXiv preprint arXiv:2502.17494}} (\bibinfo{year}{2025}).
\newblock


\bibitem[Sheng et~al\mbox{.}(2021a)]%
        {sheng2021one}
\bibfield{author}{\bibinfo{person}{Xiang-Rong Sheng}, \bibinfo{person}{Liqin Zhao}, \bibinfo{person}{Guorui Zhou}, \bibinfo{person}{Xinyao Ding}, \bibinfo{person}{Binding Dai}, \bibinfo{person}{Qiang Luo}, \bibinfo{person}{Siran Yang}, \bibinfo{person}{Jingshan Lv}, \bibinfo{person}{Chi Zhang}, \bibinfo{person}{Hongbo Deng}, {et~al\mbox{.}}} \bibinfo{year}{2021}\natexlab{a}.
\newblock \showarticletitle{One model to serve all: Star topology adaptive recommender for multi-domain ctr prediction}. In \bibinfo{booktitle}{\emph{Proceedings of the 30th ACM International Conference on Information \& Knowledge Management}}. \bibinfo{pages}{4104--4113}.
\newblock


\bibitem[Sheng et~al\mbox{.}(2021b)]%
        {sheng2021star}
\bibfield{author}{\bibinfo{person}{Xiang-Rong Sheng}, \bibinfo{person}{Liqin Zhao}, \bibinfo{person}{Guorui Zhou}, \bibinfo{person}{Xinyao Ding}, \bibinfo{person}{Binding Dai}, \bibinfo{person}{Qiang Luo}, \bibinfo{person}{Siran Yang}, \bibinfo{person}{Jingshan Lv}, \bibinfo{person}{Chi Zhang}, \bibinfo{person}{Hongbo Deng}, {et~al\mbox{.}}} \bibinfo{year}{2021}\natexlab{b}.
\newblock \showarticletitle{One model to serve all: Star topology adaptive recommender for multi-domain ctr prediction}. In \bibinfo{booktitle}{\emph{Proceedings of the 30th ACM International Conference on Information \& Knowledge Management}}. \bibinfo{pages}{4104--4113}.
\newblock


\bibitem[Shu et~al\mbox{.}(2024)]%
        {shu2024adaptive}
\bibfield{author}{\bibinfo{person}{Xiufeng Shu}, \bibinfo{person}{Ruidong Han}, \bibinfo{person}{Xiang Li}, {and} \bibinfo{person}{Wei Lin}.} \bibinfo{year}{2024}\natexlab{}.
\newblock \showarticletitle{Adaptive Utilization of Cross-scenario Information for Multi-scenario Recommendation}.
\newblock \bibinfo{journal}{\emph{arXiv preprint arXiv:2407.19727}} (\bibinfo{year}{2024}).
\newblock


\bibitem[Team et~al\mbox{.}(2026)]%
        {coreteam2026mimov2flashtechnicalreport}
\bibfield{author}{\bibinfo{person}{Core Team}, \bibinfo{person}{Bangjun Xiao}, \bibinfo{person}{Bingquan Xia}, \bibinfo{person}{Bo Yang}, \bibinfo{person}{Bofei Gao}, \bibinfo{person}{Bowen Shen}, \bibinfo{person}{Chen Zhang}, {et~al\mbox{.}}} \bibinfo{year}{2026}\natexlab{}.
\newblock \bibinfo{title}{MiMo-V2-Flash Technical Report}.
\newblock
\showeprint[arxiv]{2601.02780}~[cs.CL]
\urldef\tempurl%
\url{https://arxiv.org/abs/2601.02780}
\showURL{%
\tempurl}


\bibitem[Team et~al\mbox{.}(2025)]%
        {team2025kimi}
\bibfield{author}{\bibinfo{person}{Kimi Team}, \bibinfo{person}{Angang Du}, \bibinfo{person}{Bohong Yin}, \bibinfo{person}{Bowei Xing}, \bibinfo{person}{Bowen Qu}, \bibinfo{person}{Bowen Wang}, \bibinfo{person}{Cheng Chen}, \bibinfo{person}{Chenlin Zhang}, \bibinfo{person}{Chenzhuang Du}, \bibinfo{person}{Chu Wei}, {et~al\mbox{.}}} \bibinfo{year}{2025}\natexlab{}.
\newblock \showarticletitle{Kimi-vl technical report}.
\newblock \bibinfo{journal}{\emph{arXiv preprint arXiv:2504.07491}} (\bibinfo{year}{2025}).
\newblock


\bibitem[Tillet et~al\mbox{.}(2019)]%
        {tillet2019triton}
\bibfield{author}{\bibinfo{person}{Philippe Tillet}, \bibinfo{person}{Hsiang-Tsung Kung}, {and} \bibinfo{person}{David Cox}.} \bibinfo{year}{2019}\natexlab{}.
\newblock \showarticletitle{Triton: an intermediate language and compiler for tiled neural network computations}. In \bibinfo{booktitle}{\emph{Proceedings of the 3rd ACM SIGPLAN International Workshop on Machine Learning and Programming Languages}}. \bibinfo{pages}{10--19}.
\newblock


\bibitem[Wang et~al\mbox{.}(2021)]%
        {DCNv2}
\bibfield{author}{\bibinfo{person}{Ruoxi Wang}, \bibinfo{person}{Rakesh Shivanna}, \bibinfo{person}{Derek Cheng}, \bibinfo{person}{Sagar Jain}, \bibinfo{person}{Dong Lin}, \bibinfo{person}{Lichan Hong}, {and} \bibinfo{person}{Ed Chi}.} \bibinfo{year}{2021}\natexlab{}.
\newblock \showarticletitle{DCN V2: Improved Deep \& Cross Network and Practical Lessons for Web-scale Learning to Rank Systems}. In \bibinfo{booktitle}{\emph{Proceedings of the Web Conference 2021}} \emph{(\bibinfo{series}{WWW '21})}. \bibinfo{pages}{1785–1797}.
\newblock


\bibitem[Wang et~al\mbox{.}(2025)]%
        {wang2025mtgrboost}
\bibfield{author}{\bibinfo{person}{Yuxiang Wang}, \bibinfo{person}{Xiao Yan}, \bibinfo{person}{Chi Ma}, \bibinfo{person}{Mincong Huang}, \bibinfo{person}{Xiaoguang Li}, \bibinfo{person}{Lei Yu}, \bibinfo{person}{Chuan Liu}, \bibinfo{person}{Ruidong Han}, \bibinfo{person}{He Jiang}, \bibinfo{person}{Bin Yin}, {et~al\mbox{.}}} \bibinfo{year}{2025}\natexlab{}.
\newblock \showarticletitle{MTGRBoost: Boosting Large-scale Generative Recommendation Models in Meituan}.
\newblock \bibinfo{journal}{\emph{arXiv preprint arXiv:2505.12663}} (\bibinfo{year}{2025}).
\newblock


\bibitem[Xiaoyu et~al\mbox{.}(2025)]%
        {xiaoyu2025soft}
\bibfield{author}{\bibinfo{person}{Liu Xiaoyu}, \bibinfo{person}{Yiqing Wu}, \bibinfo{person}{Ruidong Han}, \bibinfo{person}{Fuzhen Zhuang}, \bibinfo{person}{Xiang Li}, {and} \bibinfo{person}{Wei Lin}.} \bibinfo{year}{2025}\natexlab{}.
\newblock \showarticletitle{A Soft-partitioned Semi-supervised Collaborative Transfer Learning Approach for Multi-Domain Recommendation}. In \bibinfo{booktitle}{\emph{Proceedings of the 34th ACM International Conference on Information and Knowledge Management}}. \bibinfo{pages}{5366--5370}.
\newblock


\bibitem[Yang et~al\mbox{.}(2025)]%
        {yang2025kwai}
\bibfield{author}{\bibinfo{person}{Biao Yang}, \bibinfo{person}{Bin Wen}, \bibinfo{person}{Boyang Ding}, \bibinfo{person}{Changyi Liu}, \bibinfo{person}{Chenglong Chu}, \bibinfo{person}{Chengru Song}, \bibinfo{person}{Chongling Rao}, \bibinfo{person}{Chuan Yi}, \bibinfo{person}{Da Li}, \bibinfo{person}{Dunju Zang}, {et~al\mbox{.}}} \bibinfo{year}{2025}\natexlab{}.
\newblock \showarticletitle{Kwai keye-vl 1.5 technical report}.
\newblock \bibinfo{journal}{\emph{arXiv preprint arXiv:2509.01563}} (\bibinfo{year}{2025}).
\newblock


\bibitem[Yang et~al\mbox{.}(2024)]%
        {yang2024mlora}
\bibfield{author}{\bibinfo{person}{Zhiming Yang}, \bibinfo{person}{Haining Gao}, \bibinfo{person}{Dehong Gao}, \bibinfo{person}{Luwei Yang}, \bibinfo{person}{Libin Yang}, \bibinfo{person}{Xiaoyan Cai}, \bibinfo{person}{Wei Ning}, {and} \bibinfo{person}{Guannan Zhang}.} \bibinfo{year}{2024}\natexlab{}.
\newblock \showarticletitle{Mlora: Multi-domain low-rank adaptive network for ctr prediction}. In \bibinfo{booktitle}{\emph{Proceedings of the 18th ACM Conference on Recommender Systems}}. \bibinfo{pages}{287--297}.
\newblock


\bibitem[Yuan et~al\mbox{.}(2025)]%
        {yuan2025native}
\bibfield{author}{\bibinfo{person}{Jingyang Yuan}, \bibinfo{person}{Huazuo Gao}, \bibinfo{person}{Damai Dai}, \bibinfo{person}{Junyu Luo}, \bibinfo{person}{Liang Zhao}, \bibinfo{person}{Zhengyan Zhang}, \bibinfo{person}{Zhenda Xie}, \bibinfo{person}{Yuxing Wei}, \bibinfo{person}{Lean Wang}, \bibinfo{person}{Zhiping Xiao}, {et~al\mbox{.}}} \bibinfo{year}{2025}\natexlab{}.
\newblock \showarticletitle{Native sparse attention: Hardware-aligned and natively trainable sparse attention}. In \bibinfo{booktitle}{\emph{Proceedings of the 63rd Annual Meeting of the Association for Computational Linguistics (Volume 1: Long Papers)}}. \bibinfo{pages}{23078--23097}.
\newblock


\bibitem[Zhai et~al\mbox{.}(2024)]%
        {zhai2024actions}
\bibfield{author}{\bibinfo{person}{Jiaqi Zhai}, \bibinfo{person}{Lucy Liao}, \bibinfo{person}{Xing Liu}, \bibinfo{person}{Yueming Wang}, \bibinfo{person}{Rui Li}, \bibinfo{person}{Xuan Cao}, \bibinfo{person}{Leon Gao}, \bibinfo{person}{Zhaojie Gong}, \bibinfo{person}{Fangda Gu}, \bibinfo{person}{Michael He}, {et~al\mbox{.}}} \bibinfo{year}{2024}\natexlab{}.
\newblock \showarticletitle{Actions speak louder than words: Trillion-parameter sequential transducers for generative recommendations}.
\newblock \bibinfo{journal}{\emph{arXiv preprint arXiv:2402.17152}} (\bibinfo{year}{2024}).
\newblock


\bibitem[Zhang et~al\mbox{.}(2024b)]%
        {zhang2024wukong}
\bibfield{author}{\bibinfo{person}{Buyun Zhang}, \bibinfo{person}{Liang Luo}, \bibinfo{person}{Yuxin Chen}, \bibinfo{person}{Jade Nie}, \bibinfo{person}{Xi Liu}, \bibinfo{person}{Daifeng Guo}, \bibinfo{person}{Yanli Zhao}, \bibinfo{person}{Shen Li}, \bibinfo{person}{Yuchen Hao}, \bibinfo{person}{Yantao Yao}, {et~al\mbox{.}}} \bibinfo{year}{2024}\natexlab{b}.
\newblock \showarticletitle{Wukong: Towards a scaling law for large-scale recommendation}.
\newblock \bibinfo{journal}{\emph{arXiv preprint arXiv:2403.02545}} (\bibinfo{year}{2024}).
\newblock


\bibitem[Zhang et~al\mbox{.}(2025b)]%
        {zhang2025large}
\bibfield{author}{\bibinfo{person}{Shangyu Zhang}, \bibinfo{person}{Shijie Quan}, \bibinfo{person}{Zhongren Wang}, \bibinfo{person}{Junwei Pan}, \bibinfo{person}{Tianqu Zhuang}, \bibinfo{person}{Bo Fu}, \bibinfo{person}{Yilong Sun}, \bibinfo{person}{Jieying Lin}, \bibinfo{person}{Jushuo Chen}, \bibinfo{person}{Xiaotian Li}, {et~al\mbox{.}}} \bibinfo{year}{2025}\natexlab{b}.
\newblock \showarticletitle{Large Foundation Model for Ads Recommendation}.
\newblock \bibinfo{journal}{\emph{arXiv preprint arXiv:2508.14948}} (\bibinfo{year}{2025}).
\newblock


\bibitem[Zhang et~al\mbox{.}(2024a)]%
        {zhang2024m3oe}
\bibfield{author}{\bibinfo{person}{Zijian Zhang}, \bibinfo{person}{Shuchang Liu}, \bibinfo{person}{Jiaao Yu}, \bibinfo{person}{Qingpeng Cai}, \bibinfo{person}{Xiangyu Zhao}, \bibinfo{person}{Chunxu Zhang}, \bibinfo{person}{Ziru Liu}, \bibinfo{person}{Qidong Liu}, \bibinfo{person}{Hongwei Zhao}, \bibinfo{person}{Lantao Hu}, {et~al\mbox{.}}} \bibinfo{year}{2024}\natexlab{a}.
\newblock \showarticletitle{M3oe: Multi-domain multi-task mixture-of experts recommendation framework}. In \bibinfo{booktitle}{\emph{Proceedings of the 47th International ACM SIGIR Conference on Research and Development in Information Retrieval}}. \bibinfo{pages}{893--902}.
\newblock


\bibitem[Zhang et~al\mbox{.}(2025a)]%
        {zhang2025onetrans}
\bibfield{author}{\bibinfo{person}{Zhaoqi Zhang}, \bibinfo{person}{Haolei Pei}, \bibinfo{person}{Jun Guo}, \bibinfo{person}{Tianyu Wang}, \bibinfo{person}{Yufei Feng}, \bibinfo{person}{Hui Sun}, \bibinfo{person}{Shaowei Liu}, {and} \bibinfo{person}{Aixin Sun}.} \bibinfo{year}{2025}\natexlab{a}.
\newblock \showarticletitle{OneTrans: Unified Feature Interaction and Sequence Modeling with One Transformer in Industrial Recommender}.
\newblock \bibinfo{journal}{\emph{arXiv preprint arXiv:2510.26104}} (\bibinfo{year}{2025}).
\newblock


\bibitem[Zhou et~al\mbox{.}(2025a)]%
        {zhou2025onerectechnicalreport}
\bibfield{author}{\bibinfo{person}{Guorui Zhou}, \bibinfo{person}{Jiaxin Deng}, \bibinfo{person}{Jinghao Zhang}, \bibinfo{person}{Kuo Cai}, \bibinfo{person}{Lejian Ren}, {et~al\mbox{.}}} \bibinfo{year}{2025}\natexlab{a}.
\newblock \bibinfo{title}{OneRec Technical Report}.
\newblock
\showeprint[arxiv]{2506.13695}~[cs.IR]
\urldef\tempurl%
\url{https://arxiv.org/abs/2506.13695}
\showURL{%
\tempurl}


\bibitem[Zhou et~al\mbox{.}(2025b)]%
        {zhou2025onerecv2technicalreport}
\bibfield{author}{\bibinfo{person}{Guorui Zhou}, \bibinfo{person}{Hengrui Hu}, \bibinfo{person}{Hongtao Cheng}, \bibinfo{person}{Huanjie Wang}, \bibinfo{person}{Jiaxin Deng}, \bibinfo{person}{Jinghao Zhang}, \bibinfo{person}{Kuo Cai}, {et~al\mbox{.}}} \bibinfo{year}{2025}\natexlab{b}.
\newblock \bibinfo{title}{OneRec-V2 Technical Report}.
\newblock
\showeprint[arxiv]{2508.20900}~[cs.IR]
\urldef\tempurl%
\url{https://arxiv.org/abs/2508.20900}
\showURL{%
\tempurl}


\bibitem[Zhu et~al\mbox{.}(2025)]%
        {zhu2025rankmixer}
\bibfield{author}{\bibinfo{person}{Jie Zhu}, \bibinfo{person}{Zhifang Fan}, \bibinfo{person}{Xiaoxie Zhu}, \bibinfo{person}{Yuchen Jiang}, \bibinfo{person}{Hangyu Wang}, \bibinfo{person}{Xintian Han}, \bibinfo{person}{Haoran Ding}, \bibinfo{person}{Xinmin Wang}, \bibinfo{person}{Wenlin Zhao}, \bibinfo{person}{Zhen Gong}, {et~al\mbox{.}}} \bibinfo{year}{2025}\natexlab{}.
\newblock \showarticletitle{Rankmixer: Scaling up ranking models in industrial recommenders}. In \bibinfo{booktitle}{\emph{Proceedings of the 34th ACM International Conference on Information and Knowledge Management}}. \bibinfo{pages}{6309--6316}.
\newblock


\end{thebibliography}

% \appendix
% \section{}
\end{document}